\documentclass[12pt,a4paper]{article}
\usepackage[left=1in,top=1in,right=1in,nohead]{geometry}
\usepackage{enumerate}
\usepackage{graphicx}
\usepackage{subfigure}
\usepackage{amsmath}
\usepackage{amssymb}
\usepackage{mathrsfs}
\usepackage{multirow}
\usepackage{bm}
\usepackage{color}
\usepackage{jheppub}
\usepackage{verbatim}
\usepackage{placeins}

\def\bea#1\eea{\begin{align}#1\end{align}}

\newcommand{\E}{\mathcal{E}}

\newcommand{\grad}{\nabla}

\newcommand{\arccosh}{\mathrm{arccosh}}
\newcommand{\HHRT}{$\overline{\mathrm{HHRT}}$}

\renewcommand{\Re}{\mathrm{Re}}

\def\be#1\ee{\begin{equation}#1\end{equation}}
\def\bea#1\eea{\begin{align}#1\end{align}}

\begin{document}

\title{A paucity of bulk entangling surfaces: \\ AdS wormholes with de Sitter interiors}

\author[a]{Sebastian Fischetti,}
\author[a]{Donald Marolf,}
\author[a,b]{and Aron C. Wall}

\affiliation[a]{Department of Physics \\ University of California,
Santa Barbara, Santa Barbara, CA 93106, USA}

\affiliation[b]{School of Natural Sciences, Institute for Advanced Study \\ Princeton, NJ, USA}

\emailAdd{sfischet@physics.ucsb.edu}
\emailAdd{marolf@physics.ucsb.edu}
\emailAdd{aroncwall@gmail.com}


\arxivnumber{}

\abstract{
We study and construct spacetimes, dubbed planar AdS-dS-wormholes, satisfying the null energy condition and having two asymptotically AdS boundaries connected through a (non-traversable) inflating wormhole.  As for other wormholes, it is natural to expect dual descriptions in terms of two disconnected CFTs in appropriate entangled states.  But for our cases certain expected bulk entangling surfaces used by the Hubeny-Rangamani-Takayanagi (HRT) prescription to compute CFT entropy do not exist.  In particular, no real codimension-2 extremal surface can run from one end of the wormhole to the other.   According to HRT, the mutual information between any two finite-sized subregions (one in each CFT) must then vanish at leading order in large $N$ -- though the leading-order mutual information per unit area between the two CFTs taken as wholes may be nonzero. Some planar AdS-dS-wormholes also fail to have plane-symmetric surfaces that would compute the total entropy of either CFT.  We suggest this to remain true of less-symmetric surfaces so that the HRT entropy is ill-defined and some modified prescription is required.  It may be possible to simply extend HRT or the closely-related maximin construction by a limiting procedure, though complex extremal surfaces could also play an important role.}

\maketitle

\section{Introduction}
\label{sec:intro}

The AdS/CFT correspondence~\cite{Maldacena:1997re,Witten:1998qj} offers a remarkable insight into properties of large-$N$, strongly coupled conformal field theories (CFTs): Many quantities of interest in the CFT are related to simple geometrical objects in the gravitational bulk.  Familiar examples include correlators of scalar fields with large conformal dimension that may be computed from the length of bulk geodesics~\cite{Kraus:2002iv} and Wilson loops given by the areas of bulk string worldsheets~\cite{Maldacena:1998im}.

Our interest here concerns the bulk dual of CFT entanglement entropy.  Generalizing the Ryu-Takayangi (RT) prescription \cite{Ryu:2006bv,Ryu:2006ef} to time-dependent contexts, the Hubeny-Rangamani-Takayanagi (HRT) proposal~\cite{Hubeny:2007xt} states that at leading order in $N$ the entropy of a region~$A$ of a holographic CFT is given by
\be
\label{eq:HHRT}
S(A) = \frac{\mathrm{Area}(\Xi)}{4G_N},
\ee
where $G_N$ is the bulk Newton constant and~$\Xi$ is the minimal-area (spacelike) extremal surface anchored on the set~$\partial A$.  Here we think of both $A$ and $\partial A$ as appropriate subsets of the timelike conformal boundary of an asymptotically locally AdS bulk spacetime.  Because~$\Xi$ reaches the AdS boundary, the two sides of~\eqref{eq:HHRT} are both infinite; a more meaningful equality of finite quantities follows when the two sides are properly renormalized.  As emphasized by Headrick and Takayanagi \cite{Headrick:2007km}, one should restrict attention to bulk surfaces appropriately homologous to $A$ (viewed as part of the conformal boundary).  We therefore use the term HHRT to refer both to the entire framework and to codimension-2 spacelike extremal surfaces homologous to some given $A$ (whether or not the surface has minimal area within this class).

\begin{figure}[t]
\centering
%
%
%
%
%
%
\includegraphics[page=1]{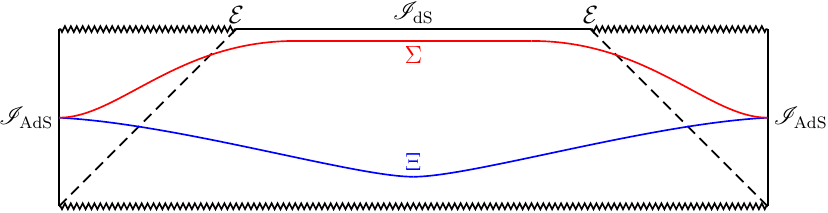}
\caption{A sample conformal diagram for an AdS-dS-wormhole.  The surface labeled~$\Xi$ (blue in color version) is a putative wormhole-spanning surface (which we will show cannot exist if the spacetime obeys the null energy condition).  The surface~$\Sigma$ (red in color version) is an achronal surface that approaches close to~$\mathscr{I}_\mathrm{dS}$ and thus has large volume element. The dashed lines indicate the boundary of the past of the~dS-like part~$\mathscr{I}_\mathrm{dS}$ of the conformal boundary. The wormhole shown has a right/left~$\mathbb{Z}_2$ reflection symmetry.  The explicit wormholes of section \ref{sec:cutpaste} will share this symmetry, though it is not needed for our general arguments.  The edges of $\mathscr{I}_\mathrm{dS}$ are marked $\E$.}
\label{fig:wormhole}
\end{figure}

The purpose of this work is to discuss HHRT for a new class of geometries, termed planar AdS-dS-wormholes.  These spacetimes describe plane-symmetric black holes with two asymptotically AdS regions connected by a wormhole that in turn contains an inflating region  -- and in particular a de Sitter-like (spacelike and smooth) region $\mathscr{I}_\mathrm{dS}$ of the conformal boundary; see figure~\ref{fig:wormhole} for an example and section \ref{sec:cutpaste} for details.  We show below that codimension-2 extremal surfaces cannot span any such wormholes, by which we mean that they cannot connect one side to the other.  It follows that HHRT predicts the leading-order large-$N$ mutual information~$I(A,B)$ to vanish between two finite-sized regions $A$ and $B$ lying on opposite conformal boundaries\footnote{We remind the reader that this mutual information can be defined in terms of the von Neumann entropies $S(A)$, $S(B)$, and $S(A\cup B)$ as
\be
\label{MI}
I(A,B) = S(A) + S(B) - S(A \cup B).
\ee}.  This is in sharp contrast to the behavior of thermofield double states studied by Hartman and Maldacena \cite{Hartman:2013qma}.

However, the leading order $I(A,B)$ is non-zero when $A$ and $B$ are the entirety of their respective boundaries since,  for that case, the empty set is also homologous to $A \cup B$.  Despite the time-dependent nature of our interior geometries, the predicted entanglement is thus similar to that of both generic entangled states (see e.g. \cite{Hartman:2013qma,Marolf:2013dba,Shenker:2013yza,Leichenauer:2014nxa} for holographic discussions) and a naive interpretation of extreme Reissner-Nordstr\"om black holes \cite{Andrade:2013rra,Leichenauer:2014nxa}.

At least when interpreted as a suitable large-torus limit of wormholes with toroidal cross sections (see section \ref{subsec:implications}), we see no inherent inconsistency in this prediction.  Indeed, further investigation of this feature may provide insights into the holographic description of inflation (see also \cite{Freivogel:2005qh,Alishahiha:2004md,Alishahiha:2005dj,Dong:2011uf}\footnote{\label{notsym} These references study time-symmetric spacetimes.  Our wormholes cannot be time-symmetric, as a moment of time-symmetry is a totally-geodesic surface.  Any wormhole-spanning minimal subsurface would thus be a wormhole-spanning extremal surface of the full spacetime.  Indeed, with planar symmetry a Raychaudhuri-equation argument like that of \cite{Farhi:1986ty} shows that no piece of $\mathscr{I}_\mathrm{dS}$ on the future boundary can lie to the future of any piece of $\mathscr{I}_\mathrm{dS}$ on the past boundary.}).  But the lack of wormhole-spanning codimension-2 surfaces makes our AdS-dS-wormholes a natural context in which to investigate possible corrections to HHRT. In particular, while the two AdS boundaries cannot be connected by any HHRT surface lying in the real Lorentz-signature spacetime, there is no obstacle to finding complex such surfaces in a complexified AdS-dS-wormhole.  Indeed, we argue below that such complex extremal surfaces exist, though we leave their detailed analysis for future work.   We remind the reader that complex saddle points often dominate the evaluation of integrals along the real axis, so that derivations of RT via saddle-point approximations to Euclidean bulk path integrals \cite{Fursaev:2006ih,Faulkner:2013yia,Hartman:2013mia,
Lewkowycz:2013nqa,Fursaev:2014tpa} naturally suggest that complex extremal surfaces be incorporated into HHRT, which would in any case require analytic continuation to make contact with the Euclidean calculation in time-dependent contexts.  See \cite{Fischetti:2014zja} for a discussion of these points, some confusions they raise, and a study of complex codimension-2 extremal surfaces in bulk duals of thermofield double states.  To leave open the question of whether \eqref{eq:HHRT} is really the CFT entropy, in what follows we will use the term ``HHRT entanglement'' to refer to the bulk quantity calculated by~\eqref{MI} using real surfaces, without implying any particular interpretation in the dual CFT.  The term ``HHRT surface'' will similarly imply the surface to be real unless explicitly stated otherwise.

We begin by constructing examples of planar AdS$_{d+1}$ dS-wormholes in section \ref{sec:cutpaste}.  We use a cut-and-paste procedure based on simpler and more familiar geometries.  The junctions where the cut-out pieces are sewn together contain distributional sources (null shells) whose stress tensors we compute.  For all $d \ge 2$ we identify cases where the result satisfies the null energy condition (NEC), both in the original spacetime from which the pieces were cut and on these null shells.

Section \ref{sec:theorem} then shows that $d \ge 2$ planar AdS$_{d+1}$ dS-wormholes obeying the NEC admit no real wormhole-spanning HHRT surfaces.  In fact, the main result is slightly more general: in any asymptotically AdS spacetime respecting the null energy condition, the light cone (boundary of the past or future) from any real codimension-2 spacelike extremal surface $\Xi$ anchored at the AdS boundary can intersect a de Sitter-like region of the conformal boundary only on a set of measure zero.  This turns out to forbid wormhole-spanning HHRT surfaces for our planar wormholes.  Regulating the geometries by allowing inflation to proceed only to a finite extent can restore theses surfaces, but their area must diverge as the regulator is removed. Either argument leads to the HHRT entanglement properties described above when $A$,~$B$ are finite-sized subsets of opposite boundaries.

The case where $A$ and $B$ are entire boundaries is discussed in section 4, where the associated HHRT surfaces are termed total entropy surfaces. Interestingly, it appears that total entropy surfaces also fail to exist in many AdS-dS-wormholes.  We show that there are no plane-symmetric total entropy surfaces in a large class of examples from section \ref{sec:cutpaste}, and we conjecture that less symmetric total entropy surfaces also fail to exist.  If so, the HHRT proposal becomes ill-defined and requires improvement.  The conceptually-simplest change would replace the HHRT surfaces with limits of families of surfaces that exist in a regulated geometry.  These limiting surfaces can be thought of as living on the conformal completion of the unregulated spacetime, so we refer to this proposal as~\HHRT.

An alternative and tempting modification, discussed in section \ref{sec:complexHM}, is the inclusion of \textit{complex} codimension-2 extremal surfaces living in complexified wormhole geometries.  Unfortunately, our cut-and-paste spacetimes are not analytic, so their complexification is far from unique.  We thus save analysis of complex surfaces in actual AdS-dS-wormholes for future work.  Instead, we analyze complex surfaces in pure de Sitter space where real surfaces again fail to exist with widely separated anchors and where we may expect a similar structure.  With help from appendix B we also note that a sum over complex geodesics accurately reproduces two-point functions of quantum fields in the de Sitter vacuum state.  Since the geodesic approximation to two-point functions shares many superficial similarities with HHRT, this provides some partial support for the idea that complex surfaces contribute to holographic entanglement for AdS-dS-wormholes.  We close with some final discussion in section \ref{sec:discussion}.

\section{Cut and Paste AdS-dS-wormholes}
\label{sec:cutpaste}

We define an AdS wormhole to be a connected solution $M$ of the Einstein equations (with a matter source respecting the null energy condition) which has two causally disconnected asymptotically (locally) AdS boundaries\footnote{With enough assumptions about the nature of these two boundaries their causal disconnection in fact follows from the null energy condition \cite{Friedman:1993ty,Galloway:1999br}.}.  AdS-dS-wormholes are those particular examples which admit a conformal extension $\overline M$ in which some piece $\mathscr{I}_\mathrm{dS}$ of the conformal boundary is, smooth, spacelike, and has diverging conformal factor.  We require $\mathscr{I}_\mathrm{dS}$ to contain an open set of the conformal boundary, and smoothness of some part the conformal boundary is taken to mean smoothness there of $\overline M$ as defined by an additional conformal factor that vanishes linearly. In the usual way these conditions imply that $M$ is asymptotically de Sitter in the region near $\mathscr{I}_\mathrm{dS}$.  Reasoning as in section 4.1 of \cite{wall:2013uza}, one may show that $\mathscr{I}_\mathrm{dS}$ must be causally inaccessible from (i.e., outside both the past and future of) any region of the AdS boundary  $\mathscr{I}_\mathrm{AdS}$.  With enough symmetry -- and in particular for planar symmetry as defined below -- this follows particularly quickly from the Raychaudhuri equation in parallel with the spherical case studied in \cite{Farhi:1986ty}; see also \cite{Firouzjahi:2002an}.  Since such spacetimes cannot be time-symmetric (see footnote \ref{notsym}), we will generally assume that $\mathscr{I}_\mathrm{dS}$ lies on the future conformal boundary as in figure \ref{fig:wormhole}.

The goal of this section is to construct simple examples of plane-symmetric AdS-dS-wormholes. This shows that such solutions exist and helps to make the discussion in the remaining sections more concrete; they are of particular use in section \ref{sec:total}.

We will build planar AdS-dS-wormholes by pasting together regions cut from more familiar spacetimes satisfying
the vacuum Einstein equations with cosmological constant, though the value of this cosmological constant will vary from region to region.  We will think of each local cosmological constant as set by a distinct extremum in the potential $V(\phi)$ of some scalar field $\phi$ which is constant in each patch.  Each junction will be a null surface, which by the Einstein equations is associated with some thin shell of matter.  For appropriate choices of parameters these null shells satisfy the null energy condition and may be interpreted as shock waves in the scalar field $\phi$.

\begin{figure}[t]
\centering
\subfigure[]{
%
%
%
%
\includegraphics[page=2]{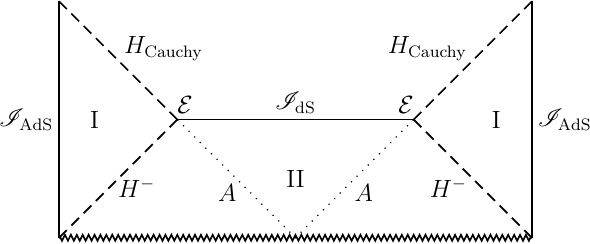}
\label{subfig:cutpastethreshold}
}\\
\subfigure[]{
%
%
%
%
\includegraphics[page=3]{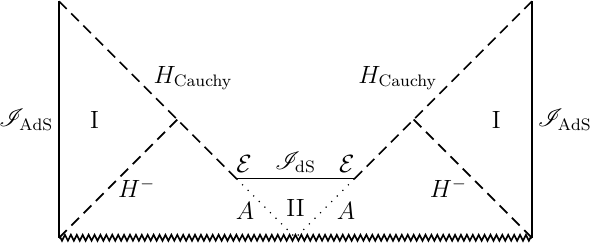}
\label{subfig:cutpastegeneral}
}
\caption{Our cut-and-paste AdS-dS-wormholes.  The two types of regions are pasted together along null shells, indicated by the dotted lines labeled~$A$, which are taken to lie along (parts of the) Killing horizons of the patches~I and ~II.  The dashed lines labeled~$H_\mathrm{Cauchy}$ are Killing horizons of patch~I and are Cauchy horizons of the full spacetime; the dashed lines labeled~$H^-$ are the past event horizons.  The two patches labeled I are isometric under a left/right reflection. \subref{subfig:cutpastethreshold}:  A case where the edges~$\E$ of $\mathscr{I}_\mathrm{dS}$ lie on the past event horizons of $\mathscr{I}_\mathrm{AdS}$.  \subref{subfig:cutpastegeneral}: A less extreme case where $\mathscr{I}_\mathrm{dS}$ lies below the past event horizon.}
\label{fig:cutpaste}
\end{figure}

We take each region to admit an additional Killing field $\xi$ beyond those involved in the planar symmetry, though the vector field $\xi$ will generally fail to be continuous across the junctions and, as a result, will not define an isometry of the full wormhole spacetime.  Our examples will be built from three such patches, but we impose a~$\mathbb{Z}_2$ reflection symmetry exchanging the ends of the wormhole so that these regions are of only two distinct types (called I and II, see figure~\ref{fig:cutpaste}).

\begin{figure}[t]
\centering
%
%
%
\includegraphics[page=4]{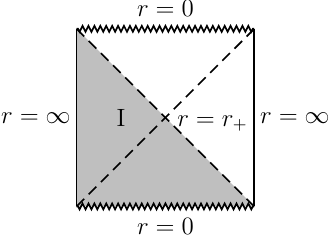}
\hspace{1cm}
%
%
%
\includegraphics[page=5]{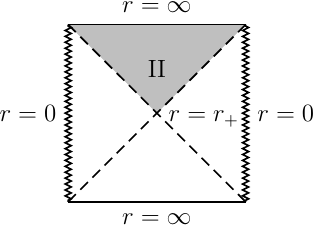}
\caption{Conformal diagrams for the spacetimes from which we cut our (shaded) regions~I and~II.  The dashed lines on both diagrams correspond to the Killing horizons at~$r = r_+$.  For simplicity we do not show the relative bending between the singularity and boundary.}
\label{fig:pieces}
\end{figure}

Region I will be the part of the familiar planar AdS$_{d+1}$-Schwarzschild black hole (or BTZ for~$d = 2$) lying to the past of one AdS boundary, while region II is (part of) an analytic continuation of the planar AdS$_{d+1}$-Schwarzschild black hole to positive effective cosmological constant (studied in~\cite{Das:2013mfa}; see \eqref{eq:patchmetric} and \eqref{eqs:fs} below).  The conformal diagrams of these spacetimes and the indicated regions are shown in figure~\ref{fig:pieces}.  Each patch extends to the relevant part of the future and/or past Killing horizon.

The junctions are two copies of a single null shell (drawn as dotted lines and both labeled~$A$ in the figure) which lie on parts of the would-be Killing horizons of~$\xi$.  Note that our wormhole has Cauchy horizons~$H_\mathrm{Cauchy}$ along other pieces of the would-be Killing horizons.  In analogy with the Reissner-Nordstr\"om case \cite{Poisson:1990eh,Ori:1991zz,Dafermos:2003wr}, we expect our Cauchy horizons to be unstable to forming null singularities.  They should thus be considered an artefact of our cut-and-paste construction.

We also introduce a coordinate~$r$ defined at each point by the scale factor of the corresponding plane of symmetry, and which must be continuous across each shell.  This requires the black hole horizon in patch I to have the same ``radius'' $r_+$ as the de Sitter horizon in patch II, though the effective cosmological constant (parametrized by the associated length scales $\ell_I, \ell_{II}$) and black hole mass-density may differ.  As noted above, one may think of the associated jumps as modeling gravity coupled to a scalar field whose potential has both AdS and dS extrema.

In both patches the metric thus takes the form
\be
\label{eq:patchmetric}
ds^2_n = -f_n(r) dt_n^2 + \frac{dr^2}{f_n(r)} + r^2 \, d\vec{x}_{d-1}^2,
\ee
where $n = $~I,~II, each of the~$f_n$ have a zero at the same value~$r = r_+$, and the coordinates~$t_n$ will generally differ from patch to patch.  In particular, we take
\begin{subequations}
\label{eqs:fs}
\bea
f_I(r) &= \frac{r^2}{\ell_I^2}\left(1-\left(\frac{r_+}{r}\right)^d\right), \\
f_{II}(r) &= -\frac{r^2}{\ell_{II}^2}\left(1-\left(\frac{r_+}{r}\right)^d\right).
\eea
\end{subequations}
In regions~I and~II, we have~$0 < r < \infty$ and~$r_+ < r < \infty$ respectively, as shown in figure~\ref{fig:pieces}.

Assembling these patches as in figure \ref{fig:cutpaste} yields a planar AdS$_{d+1}$ dS-wormhole.  But the result is far from unique, as we must specify the manner in which each pair of regions is sewn together at the relevant junction.  In our context, it is convenient to do so using Eddington-Finkelstein coordinates
\be
\label{eq:EF}
du_n = dt_n - \frac{dr}{f_n(r)}, \quad dv_n = dt_n + \frac{dr}{f_n(r)}.
\ee
Recall that the past event horizon in region I is $v_I = - \infty$, with $v_I$ running from $\infty$ to $-\infty$ below this horizon and then again from $-\infty$ to $\infty$ above.  We sew patch II to the part of patch I below the past horizon using
\be
v_{II} = \frac{1}{\kappa_{II}} \, g_A\left(\kappa_I v_I\right),
\ee
where~$g_A(x)$ is an arbitrary continuous monotonic function. The fact that we placed the boundaries of our regions at Killing horizons means that the induced metric is continuous across the junction for any~$g_A$.  To construct figure~\ref{subfig:cutpastethreshold}, we choose $g_A$ to map~$(-\infty,\infty) \mapsto (-\infty,\infty)$, and in particular take $g_A(-\infty) = -\infty$.  This takes the two edges~$\E$ of $\mathscr{I}_\mathrm{dS}$ to lie precisely on the past event horizons~$H^-$ of $\mathscr{I}_\mathrm{AdS}$, as shown in figure~\ref{subfig:cutpastethreshold}.  Recall from earlier that, as in \cite{Farhi:1986ty}, the null energy condition prevents $\mathscr{I}_\mathrm{dS}$ from being to the future of any point of $\mathscr{I}_\mathrm{AdS}$, so this current case is a threshold case.  We may move $\mathscr{I}_\mathrm{dS}$ lower (as in figure~\ref{subfig:cutpastegeneral}), but no higher.

That the spacetimes of figure~\ref{fig:cutpaste} obey the NEC is easily verified by calculating the stress tensor of our shells.  The key quantities are their energy density $\mu$ and pressure $p$.  We wish to find examples satisfying $\mu + p \ge 0$; this condition is equivalent to the NEC in our context.  The computations are described in appendix~\ref{app:shells}.  For the spacetime of figure~\ref{subfig:cutpastethreshold}, we take~$g_A(x) = \beta x$; then using~\eqref{eq:BT} the condition~$\mu_A + p_A \geq 0$ becomes equivalent to
\be
\kappa_{II} \geq \beta \kappa_I \left[1-\frac{1+\beta}{d-1} \, \kappa_I r_+ \right].
\ee
Choosing, for instance,
\be
\label{eqs:params}
\quad \kappa_I r_+ = \frac{1}{4}, \quad \kappa_{II} r_+ = \frac{1}{4}\left(1-\frac{1}{2(d-1)}\right), \quad \beta = 1
\ee
yields $\mu_A + p_A = 0$ for all $d \geq 2$, giving a patched AdS-dS-wormhole that saturates the NEC everywhere.

To construct figure~\ref{subfig:cutpastegeneral}, we instead set $g_A(x) = \ln(e^{dx}-e^{d\kappa_I v_0})$ and take the domain of $v_I$ to be~$(v_0,\infty)$; this places the edges~$\E$ of~$\mathscr{I}_\mathrm{dS}$ at a finite advanced time~$v_I = v_0$ and yields
\be
\mu_A + p_A = \frac{1}{8\pi G_N r_+}\left[\frac{2d}{e^{d\kappa_I (v_I-v_0)}-1} + 2 + \left(1-e^{-d\kappa_I (v_I-v_0)}\right)\left(1-\frac{1}{d}\right)\right],
\ee
which is positive\footnote{One may ask if in analogy with the threshold case there exists some choice of parameters that saturates the NEC; that is, for arbitrary finite~$v_0$, is there a choice of~$\kappa_I > 0$,~$\kappa_{II} > 0$, and smooth monotonic~$g_A(x)$ with domain~$(v_0,\infty)$ that sets~$\mu_A + p_A = 0$?  The answer is no: using~\eqref{eq:BT} the condition~$\mu_A + p_A = 0$ becomes a differential equation for~$g_A$, whose only solutions do not obey the monotonicity requirement.} for all~$v_I > v_0$ and~$d \geq 2$.

\begin{figure}[t]
\centering
\subfigure[]{
%
%
%
%
\includegraphics[page=6]{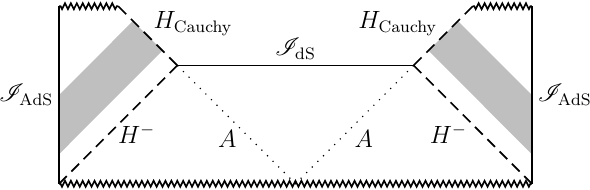}
\label{subfig:Vaidyashell}
} \\
\subfigure[]{
%
%
%
%
\includegraphics[page=7]{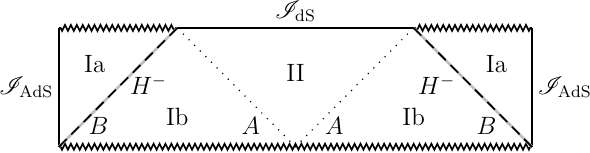}
\label{subfig:doublepatched}
}
\caption{Firing in matter from the AdS boundaries modifies the cut-and-paste wormholes of figure~\ref{fig:cutpaste}. The spacetimes shown are based on figure~\ref{subfig:cutpastethreshold}, though corresponding results also hold for figure~\ref{subfig:cutpastegeneral}.  \subref{subfig:Vaidyashell}: Patch~I is replaced by an AdS-Vaidya metric representing pressureless null dust (shaded) falling in from $\mathscr{I}_\mathrm{AdS}$. This adds a future singularity that cuts off part of the Cauchy horizon $H_\mathrm{Cauchy}$.  \subref{subfig:doublepatched}: One can remove the Cauchy horizon completely by firing in a thin null shell $B$ (light gray lines beneath dashed lines) along~$H^-$.  The shell further divides region I into subregions Ia and Ib on either side.  This shell cannot be pressureless (see footnote \ref{note:pressure}) and is not a simple limit of the Vaidya case shown at top.}
\label{fig:Vaidyapatching}
\end{figure}

As noted above, our cut-and-paste construction led to a Cauchy horizon~$H_\mathrm{Cauchy}$.  While not a problem for our later discussion and likely unstable, we nevertheless mention that it is easy to shrink this horizon or even remove it entirely by including further simple matter sources.  For example, one can fire null dust (obeying the null energy condition) from the AdS boundary, as shown in figure~\ref{subfig:Vaidyashell}.  This replaces the pure AdS-Schwarzschild metric in the part of patch~I above~$H^-$ with an ingoing planar AdS-Vaidya metric (i.e., the ingoing planar AdS analogue of \cite{Vaidya:1999zz,Vaidya:1951zz}) of the form
\be
\label{eq:AdSVaidya}
ds^2_I = -f_I(r,v_I) dv_I^2 + 2 \, dv_I \, dr + r^2 \, d\vec{x}_{d-1}^2, \mbox{ where } f_I(r,v_I) = \frac{r^2}{\ell_I^2}\left(1 - \frac{\tilde r^d(v_I)}{r^d} \right),
\ee
where $\tilde r(v_I)$ is an arbitrary function satisfying~$\tilde r'(v_I) \geq 0$ and~$\tilde r(-\infty) = r_+$.

In principle, the Cauchy horizon can be made to disappear entirely by firing in a thin null shell along~$H^-$ itself. The spacetime then becomes the one shown in figure~\ref{subfig:doublepatched}.  Furthermore, as the new null shell runs along a would-be Killing horizon, each of the (now five) patches~Ia,~Ib, and~II still admits a timelike Killing field~$\xi$.  In appendix~\ref{app:shells} we show by explicit construction that the resulting spacetime does indeed obey the null energy condition, though since the new null shell is not pressureless, it is not in any simple sense a limiting case of the Vaidya spacetime\footnote{\label{note:pressure} That the shell cannot be pressureless follows from the fact that the pressure of the shell is a measure of the discontinuity in the acceleration of its generators across it~\cite{Poisson}.  Since its generators are future inextendible (extendible) with respect to patch~Ia~(Ib), this discontinuity must be nonzero and the shell pressure cannot vanish everywhere.}.

\section{No localized HHRT entanglement}
\label{sec:theorem}

We turn now to HHRT surfaces and entanglement.  The goal of this section is to show for all $d \ge 2$ that, according to HHRT,  planar AdS$_{d+1}$-dS-wormholes describe states defined on two CFTs in which the CFTs are jointly pure at leading order in large $N$ but which have vanishing leading-order mutual information between finite-sized subregions of opposing boundaries. The leading-order purity of the total state is straightforward:  the pair of AdS boundaries taken together is homologous to the empty set so that the total leading-order entropy vanishes.  And the mutual information \eqref{MI} will vanish between finite-sized subregions $A$, $B$ if $S(A \cup B) = S(A) + S(B)$.

We show below that AdS-dS-wormholes have no wormhole-spanning codimension-2 extremal surfaces.  So when $A$, $B$ are finite-sized subregions of opposite boundaries, every extremal surface anchored on $\partial(A \cup B)$ is in fact the union of two disconnected surfaces -- one anchored on $\partial A$ and the other on $\partial B$. A naive application of HHRT thus yields $S(A \cup B) = S(A) + S(B)$ and $I(A,B)=0$.  In order to take a bit more care, we also consider regulated versions of our spacetimes where wormhole-spanning surfaces do exist and show that this behavior is reproduced in the limit where the regulator is removed.

The arguments of this section do not in fact require the full planar symmetry; it is enough to have the translation subgroup.  We refer to this as planar-translation symmetry in order to distinguish it from full planar symmetry.

\subsection{No wormhole-spanning extremal surfaces}
\label{subsec:van}

We first show that the intersection of de Sitter-like regions of the conformal boundary with the light cone (boundary of the past or future) from any real codimension-2 extremal surface $\Xi$ must have measure zero when the  only boundaries of $\Xi$ lie at the AdS boundaries.  We will refer to this latter property saying that $\Xi$ is
anchored at the AdS boundary.   As for $\mathscr{I}_\mathrm{dS}$ above, we define de Sitter-like regions of the conformal boundary to be those that are smooth and spacelike with divergent conformal factor.  We assume the spacetime to satisfy the null convergence condition $R_{ab} k^a k^a \ge 0$, which holds for solutions of the Einstein-Hilbert equations of motion for gravity coupled to matter that respects the null energy condition. The argument is closely related to the methods of \cite{Wall:2012uf}.     Where not specified, we will use the conventions and definitions of~\cite{Wald:1984}.

To begin, consider a real codimension-2 extremal surface $\Xi$ anchored at the AdS boundary whose light cone intersects a de Sitter-like region of the conformal boundary.  Since the only boundary of $\Xi$ lies at the AdS boundary, and since any extremal surface intersects the AdS boundary orthogonally, the light cone of $\Xi$ is generated by a congruence of null geodesics fired orthogonally from~$\Xi$.  Furthermore, since $\Xi$ is extremal and codimension-2, the expansion of this congruence vanishes at $\Xi$.  No new generators can join the light cone as one moves away from $\Xi$,  and the null convergence condition implies that the expansion can only decrease.  Thus, just as in the proof of the Hawking area theorem \cite{Hawking:1971tu}, the area of the light cone can only decrease as one moves away from $\Xi.$

On the other hand, any piece of this light cone which intersects a de Sitter-like region of the conformal boundary on a set of non-zero measure has infinite area.  If $\Xi$ has finite area this immediately implies that the intersection must have measure zero.  If $\Xi$ has infinite area (as in the case of interest), the same conclusion is reached by considering a compact set of null geodesics in our congruence that reach the de Sitter-like infinity; they must have been fired from a compact subset of $\Xi$ with finite area.   And if all compact subsets have zero measure then the total measure of the intersection must vanish as well.  This argument assumes the light cone to be piecewise $C^2$ in parallel with Hawking's original derivation \cite{Hawking:1971tu} of the area theorem, but we expect that this assumption can be dropped using the methods of \cite{Chrusciel:2000cu}.

One may use the above result to exclude wormhole-spanning HHRT surfaces in an AdS-dS-wormhole with an
an everywhere-spacelike freely-acting ${\mathbb R}^{d-1}$ translation symmetry (which we call planar-translation symmetry), or in any quotient of such a spacetime by any subgroup of these translations.  For such a translation-planar AdS-dS-wormhole $M$, it is natural to consider conformal extensions $\overline M$ containing $\mathscr{I}_\mathrm{dS}$ for which the relevant conformal factor and thus $\overline M$ are also invariant under this planar-translation symmetry.  This means that $\overline M$ cannot be compact, but we will choose conformal extensions that become so under any quotient by a discrete translation subgroup group that takes ${\mathbb R}^{d-1}$ to the torus ${T}^{d-1}$.

The planar-translation symmetry implies that any wormhole-spanning extremal surface $\Xi$ must pass through the region to the past of $\mathscr{I}_\mathrm{dS}$, see figure~\ref{fig:wormhole}. But the light cone of $\Xi$ can expand only with finite speed in the conformally extended spacetime $\overline M$, while $\overline M$ remains infinite in the planar directions. Thus the part of $\mathscr{I}_\mathrm{dS}$ to the future of $\Xi$ can be of only finite extent in the planar directions.  Since $\mathscr{I}_\mathrm{dS}$ is invariant under the full infinite planar-translation symmetry, the future light cone of $\Xi$ (i.e., the boundary of its future) must intersect $\mathscr{I}_\mathrm{dS}$ along some surface that spans $\mathscr{I}_\mathrm{dS}$ from one end to the other.  And since $\mathscr{I}_\mathrm{dS}$ is non-trivial, the measure of this intersection is non-zero.  This contradicts the result above and shows that $\Xi$ cannot exist.  It also follows that wormhole-spanning extremal surfaces cannot exist in any quotient as they would then lift to a wormhole-spanning extremal surface in the covering spacetime $M$.

We note that this same result can be derived directly using the maximin prescription of~\cite{Wall:2012uf}  (which was shown to be equivalent to HHRT in certain contexts).  The maximin construction considers all achronal surfaces $\Sigma$ satisfying appropriate boundary conditions, such as the one shown in figure \ref{fig:wormhole}.  One then finds the minimal surface on each $\Sigma$ and then maximizes the area of this surface over all $\Sigma$.  So the area of the maximin surface is bounded below by the area of the minimal surface on any given $\Sigma$.  Since  $\mathscr{I}_\mathrm{dS}$ is outside the light cone of any point on any AdS boundary,  we may choose $\Sigma$ to lie arbitrarily close to $\mathscr{I}_\mathrm{dS}$ over a finite portion of its length as shown in figure \ref{fig:wormhole}.  In the limit where $\Sigma$ approaches $\mathscr{I}_\mathrm{dS}$ in this way the area of the minimal wormhole-spanning surface on $\Sigma$ grows without bound.  We see that the area of any maximin surface must be infinite, and that no actual maximin surface can exist in $M$.  This argument works directly in both translation-planar spacetimes and their quotients.

On the other hand, there is no obstruction to having extremal codimension-2 surfaces outside the horizon.  Indeed, we may take the exterior regions of our wormhole to be just planar AdS-Schwarzschild in which extremal surfaces have been extensively studied~(e.g. in \cite{Ryu:2006bv,Ryu:2006ef}).  Considering finite-sized subregions $A$ and $B$ of opposite boundaries, the lack of wormhole-spanning extremal surfaces means that, when the translation-symmetry is non-compact, a naive application of HHRT finds the minimal area surface computing $S(A \cup B)$ to be disconnected, with each connected component giving just $S(A)$ or $S(B)$ separately\footnote{The reader may note that $A \cup B$ is homologous to $\bar A \cup \bar B$ where $\bar A, \bar B$ are the complements of $A,B$ within their respective boundaries. As a result, there are also disconnected surfaces with each piece separately homologous to $\bar A, \bar B$.  But when the translation symmetry is non-compact and $A, B$ are finite-sized, these latter surfaces will have infinite area and do not contribute.  For toroidal wormholes, they will again fail to contribute when $A, B$ are sufficiently small but make the leading-order $I(A,B)$ non-zero for large enough $A,B$.}. In other words, our result implies $S(A \cup B) \approx S(A) + S(B)$ so that $I(A,B) \approx 0$, where $\approx$ denotes equality at leading order in large $N$.

We now pause to evaluate this conclusion more carefully.  In particular, we consider regulated versions of our AdS-dS-wormholes in which inflation proceeds only for a finite time before the wormhole recollapses to a singularity.  Simplified models of such spacetimes are constructed and studied in detail in appendix \ref{app:regulated}.  Removing $\mathscr{I}_\mathrm{dS}$ in this way allows wormhole-spanning HHRT surfaces to exist.  Indeed, the arguments of \cite{Wall:2012uf} tell us that they do, and that they coincide with maximin surfaces\footnote{\label{sing} The theorems in \cite{Wall:2012uf} address Kasner-like singularities.  The singularities of our regulated wormholes are naturally either of Kasner-like or of the `big crunch' form where all directions shrink to zero size.  Since all surfaces near the big crunch are small, it is manifest that the maximization step of the maximin procedure keeps one well away from such singularities.  It is thus even easier to apply the arguments of \cite{Wall:2012uf} in this case than for Kasner-like singularities.}.

The maximization step in the maximin procedure suggests that wormhole-spanning extremal surfaces lie near the surface of maximal inflation in the regulated wormhole.  More precisely, we argue in appendix \ref{app:regulated} that at late times they approach a surface of maximal effective scale factor in behavior analogous to that found by Hartman and Maldacena in AdS-Schwarzschild \cite{Hartman:2013qma}.    This surface recedes to $\mathscr{I}_\mathrm{dS}$ and becomes of infinite size in any limit where our regulator is removed.  In contrast, the area of disconnected surfaces that lie outside the horizon will remain finite as the regulator is removed.  So, as above, when the translation symmetry group is appropriately non-compact, HHRT again predicts $S(A \cup B) \approx S(A) + S(B)$ for AdS-dS-wormholes and $I(A,B)\approx0$.

\section{No total entropy surfaces in $M$, but finite total entropy}
\label{sec:total}

We have seen that planar AdS-dS-wormholes have vanishing HHRT entropy between finite-sized subregions of opposite boundaries.  This raises the question of taking $A$ and $B$ to be (opposite) boundaries in their entirety.  Since $A \cup B$ is then homologous to the empty set, HHRT finds $S(A \cup B) =0$ and $I(A,B) = S(A) + S(B) = 2S(A)$.  But it remains to compute $S(A)$ by finding the associated HHRT surfaces.  Such (putative) surfaces are called total entropy surfaces below.

For a broad class of planar AdS-dS-wormholes from section \ref{sec:cutpaste}, section \ref{subsec:nomaximin} will demonstrate that plane-symmetric total entropy surfaces do not exist in the physical spacetime $M$.  This argument uses the full planar symmetry and not just the translation subgroup, though corresponding results follow immediately for toroidal quotients.  We conjecture that less-symmetric total entropy surfaces also fail to exist and that the HHRT entropy is ill-defined.  More complicated examples similarly suggest that a strict application of HHRT gives physically incorrect results even when a total entropy surface exists in $M$.

Consideration of regulated spacetimes in section \ref{subsec:entropy} nevertheless argues that HHRT be extended to assign a finite entropy to each boundary of our AdS-dS-wormholes.  The non-zero entropy implies a positive mutual information between the two boundaries.  We also locate an effective HHRT surface lying in the conformal boundary at the edge of $\mathscr{I}_\mathrm{dS}$.  The implications for entanglement are summarized in section \ref{subsec:implications}.

\subsection{Planar wormholes without planar total entropy surfaces}
\label{subsec:nomaximin}

The example wormholes of section \ref{sec:cutpaste} have full planar symmetry, including reflections as well as translations in each (spacelike) planar direction.  This implies that our wormholes admit unique (future-directed) left- and right-moving null congruences orthogonal to every orbit of the planar symmetry group; i.e., whose velocity field has only $r,t$ components.  Since a codimension-2 surface is extremal if and only if the expansion vanishes at the surface for each of the two orthogonal null congruences,  plane-symmetric total entropy surfaces arise only when the left- and right-moving congruences define zero-expansion surfaces ($\theta_L=0$,~$\theta_R=0$) that intersect.

One can certainly find AdS-dS-wormholes where this intersection exists.  For example, this occurs when the wormhole exterior is precisely AdS-Schwarzschild up to and including the bifurcation surface.  The left- and right-moving AdS-Schwarzschild Killing horizons have respectively $\theta_L=0$,~$\theta_R=0$ and intersect at a total entropy surface (i.e. the bifurcation surface).  But there are other choices where the zero-expansion surfaces do not intersect.

\begin{figure}[t]
\centering
\subfigure[]{
%
%
%
%
\includegraphics[page=8]{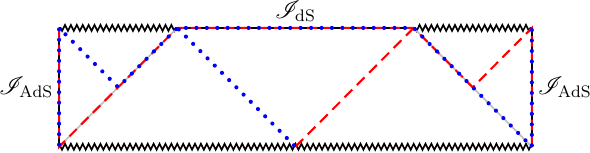}
\label{subfig:thetazerodoublepatched}
} \\
\subfigure[]{
%
%
%
%
\includegraphics[page=9]{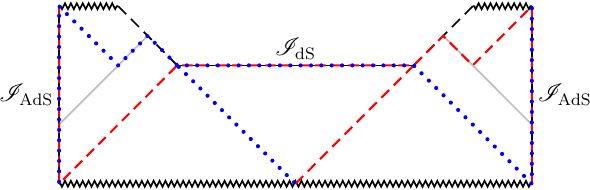}
\label{subfig:thetazeroVaidya}
}
\caption{Surfaces of $\theta_R=0$ (dashed lines; red in color version) and $\theta_L=0$ (dotted lines; blue in color version) for the AdS-dS-wormholes shown in figure~\ref{fig:Vaidyapatching}.  Note that since affine parameters diverge at $\mathscr{I}_\mathrm{dS}$ and $\mathscr{I}_\mathrm{AdS}$, the Raychaudhuri equation guarantees that $\theta_R$, $\theta_L$ both vanish on these surfaces.  We take the ingoing matter to consist of null shells (solid gray lines).  \subref{subfig:thetazerodoublepatched}: The spacetime of figure~\ref{subfig:doublepatched}.
Null shells with non-zero pressure are fired in along the past horizons of $\mathscr{I}_\mathrm{AdS}$; this fine-tuning leads the~$\theta_R = 0$,~$\theta_L = 0$ surfaces to overlap along portions of these past horizons.  \subref{subfig:thetazeroVaidya}: When the incoming shells are displaced to the future the surfaces~$\theta_R = 0$,~$\theta_L = 0$ no longer intersect in~$M$ and total entropy surfaces do not exist in~$M$. Here the shell may be chosen pressureless so that this case is a simple limit of figure \ref{subfig:Vaidyashell}.  A version in which this new null shell is smoothed out is shown in figure \ref{fig:thetazeroVaidya}.}
\label{fig:thetazeroshells}
\end{figure}

For planar congruences in planar spacetimes the sign of the expansion is positive when $r$ increases along the congruence and negative when it decreases.  So it is straightforward to draw $\theta_L=0$,~$\theta_R=0$ contours for the simple cases shown in figures \ref{subfig:cutpastethreshold} and~\ref{subfig:doublepatched} in which the matter consists only of thin shells.   The results are shown in figure~\ref{fig:thetazeroshells}.  Since the expansions are generally not continuous at the junctions, in most cases what we have actually drawn is the boundary between the region of positive expansion (below the indicated lines) and the region of negative expansion (above the indicated lines)\footnote{The exception occurs at shell A, where on either side the congruence along this shell has positive expansion that vanishes as the shell is approached.}.

\begin{figure}[t]
\centering
%
\includegraphics[page=10]{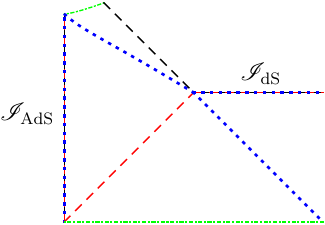}
\caption{Surfaces of zero expansion in the~$d = 2$ AdS-Vaidya dS-wormhole with the mass function~\eqref{eq:Vaidyamu}.  Conventions are the same as in figure~\ref{fig:thetazeroshells}, except that singularities are now drawn as solid lines (green in color version).  Note that we only show the left half of the spacetime, and that only surfaces to the future of the past horizon~$H^-$ have been calculated and plotted explicitly; the past singularity and~$\mathscr{I}_\mathrm{dS}$ have been drawn as straight lines by hand.  The Cauchy horizon intersects the singularity at advanced time~$v_0 = -\ell_I \ln \sqrt{3}$. As in figure \ref{fig:thetazeroshells}, this Cauchy horizon can also be removed by adding a null shell along $H^-$.}
\label{fig:thetazeroVaidya}
\end{figure}

When the matter shells enter along the past event horizons of $\mathscr{I}_\mathrm{AdS}$ (as in figure~\ref{subfig:thetazerodoublepatched}) we find that $\theta_L=0$, $\theta_R=0$ surfaces coincide over a finite piece of these horizons near $\mathscr{I}_\mathrm{dS}$.  But this is an artefact of the associated fine tuning.  Taking the shell to enter later (as in figure~\ref{subfig:thetazeroVaidya}) displaces the outgoing zero-expansion surface toward the future so that the two surfaces no longer intersect in the physical spacetime $M$.   For appropriate choices, this remains true when we smooth out the thin shell by passing to the Vaiya wormhole shown in figure~\ref{subfig:Vaidyashell}; see figure~\ref{fig:thetazeroVaidya} for an explicit example which takes~$d = 2$ and
\be
\label{eq:Vaidyamu}
\tilde r(v_I) = r_+ \sqrt{5+4\tanh(v_I/\ell_I)},
\ee
with~$\tilde r(v)$ defined as in~\eqref{eq:AdSVaidya}.  The Cauchy horizons in these examples should be unstable and non-generic as described in section \ref{sec:cutpaste}, though we see no reason that such instabilities should restore the missing total entropy surfaces.

In such cases there can be no planar total entropy surface.  The same is clearly true of toroidal quotients.  We expect that less-symmetric total entropy surfaces fail to exist as well\footnote{In the past domain of dependence of $\mathscr{I}_\mathrm{dS}$,  extremal surfaces that extend in the planar directions will tend to bend toward the singularities.  But closed surfaces in $M$ will have points (locally) ``closest'' to the singularity.  So one need only exclude extremal surfaces from other regions of the conformal diagram.}.

\subsection{Regulated wormholes}
\label{subsec:entropy}

The lack of total entropy surfaces in these cases renders the HHRT entropy of either boundary ill-defined.  So this prescription clearly requires modification. When wormhole-spanning extremal surfaces did not exist in section \ref{sec:theorem}, we argued that they could equivalently be assigned infinite entropy.  But taking the $x^i$ coordinates periodic turns each boundary into a finite torus (at each time).  So since the bulk clearly has finite energy, it would be physically incorrect to assign infinite entropy to either CFT.  Some other resolution is needed.

Useful insight can again be obtained by considering the regulated and smoothed-out wormholes of section~\ref{sec:theorem}; the key point is again that they inflate only to a finite extent before recollapsing to a singularity. Thus all desired extremal surfaces will exist (see footnote \ref{sing}). Furthermore, in these regulated spacetimes, theorem 16 of \cite{Wall:2012uf} guarantees total entropy surfaces to have smaller area $A_{TE}^\mathrm{reg}$ than the area~$A_\mathrm{bif}$ of the smallest bifurcation surface of either the right or left event horizon.  So holding~$A_\mathrm{bif}$ fixed as the regulator is removed gives a regulator-independent upper bound on $A^\mathrm{reg}_{TE}$.  In particular, since the Hawking area theorem guarantees the late-time area of the event horizon to be even larger, the bound on $A_{TE}^\mathrm{reg}$ is consistent with the expected CFT density of states at the given energy.

This suggests that $A^\mathrm{reg}_{TE}$ may approach some limit $A_{TE}^\mathrm{lim}$ as the regulator is removed.  Using $A_{TE}^\mathrm{lim}$ to calculate entropy for AdS-dS-wormholes would be a simple extension of HHRT that we christen \HHRT, though we will not study convergence of this limit in any detail.  However, we mention that more complicated variations on the above examples suggest that the original HHRT prescription can assign the wrong entropy even when a total entropy surface does exist.  For example, we could modify the spacetimes of figure \ref{fig:Vaidyapatching} by adding a further AdS-Schwarzschild region with unmolested bifurcation surfaces that introduce new extremal surfaces.  If the area $A_\mathrm{new}$ of this new surfaces exceeds the above $A_{TE}^\mathrm{lim}$, then HHRT will use a smaller surface to compute our entropy in any regulated spacetime.  Strict use of HHRT would then predict the entropy to be discontinuous as the regulator is removed while by construction \HHRT~gives a continuous result.  And, as above, in many cases the strict HHRT result will give~$A_\mathrm{new} > A_\mathrm{bif}$ which will often conflict with the CFT density of states as set by the total energy (while consistency of \HHRT~is guaranteed).

It also is useful to discuss the above limit in terms of the maximin prescription of \cite{Wall:2012uf}, which is equivalent to HHRT in our regulated context (see again footnote \ref{sing}).  We once more recall that a maximin surface is constructed by first studying all achronal surfaces $\Sigma$, identifying the minimal surface on each, and maximizing the associated area over all $\Sigma$.  Now, the proposal that AdS-dS-wormholes have no maximin total entropy surface in the physical spacetime $M$ would mean that this final maximum does not exist.  But since~\cite{Wall:2012uf} guarantees that the minimal surface on any~$\Sigma$ has area smaller than~$A_\mathrm{bif}$, we may still discuss the least upper bound $A_{TE}^\mathrm{lub}$ of the areas over all achronal surfaces.  And for the above toroidal wormholes this $A_{TE}^\mathrm{lub}$ must be finite, as it is also bounded above by the area of the horizon bifurcation surface.

For simplicity, let us suppose that $\mathscr{I}_\mathrm{dS}$ lies in the future conformal boundary.  Then our regulator deforms the AdS-dS-wormhole only in the far future.  In particular, any achronal surface in the AdS-dS-wormhole is also an achronal surface in regulated wormholes with sufficient amounts of inflation.  It thus persists as the regulator is removed and gives a lower bound on the limit $A_{TE}^\mathrm{lim}$.  It follows that $A_{TE}^\mathrm{lim}$ is at least $A_{TE}^\mathrm{lub}$.

On the other hand, suppose that some regulated spacetime had $A^\mathrm{reg}_{TE}$ greater than $A_{TE}^\mathrm{lub}$.  Then the achronal surface containing this maximin surface can have no counterpart in the unregulated wormhole.  One thus expects to be able to use regulators where $A^\mathrm{reg}_{TE}$ converges precisely to $A_{TE}^\mathrm{lub}$ as the regulator is removed; i.e., for which $A_{TE}^\mathrm{lim} = A_{TE}^\mathrm{lub}$.  It is therefore natural to extend the maximin prescription to our AdS-dS-wormholes by assigning entropy $A_{TE}^\mathrm{lub}/4G_N$ to each CFT and to term the associated scheme $\overline{\mathrm{maximin}}$ regardless of the conditions under which this coincides with a limit of $A^\mathrm{reg}_{TE}$.  We remark that for the spacetimes of figure \ref{fig:Vaidyapatching} this $A_{TE}^\mathrm{lub}$ is precisely the area $A_{H^-} = r_+^{d-1} V_{d-1}$ of the past horizon~$H^-$ of $\mathscr{I}_\mathrm{AdS}$.  Here $V_{d-1}$ is the coordinate volume of the $x^i$ directions and we have already argued $A_{TE}^\mathrm{lub} \le A_{H^-}$.  Since achronal surfaces close to the future boundary have $r > r_+ - \epsilon$ everywhere for any $\epsilon > 0$ we also have $A_{TE}^\mathrm{lub} \ge A_{H^-}$ and thus $A_{TE}^\mathrm{lub} = A_{H^-}$.

A useful feature of the original HHRT framework was that it associated the entropy calculation with a specific surface in the bulk.  In particular, we recall that this observation has led to proposals \cite{Czech:2012bh,Wall:2012uf} for the bulk region dual to subregions of a CFT; see also \cite{Bousso:2012sj,Hubeny:2012wa}.  It would thus be nice to locate a surface to which we can assign area $A_{TE}^\mathrm{lub}$.

There is of course no natural candidate in the physical unregulated spacetime $M$.  But we can ask if the total entropy surfaces of the regulated spacetimes converge in any sense to a surface in the conformal extension $\overline M$.  Note that, since our regulator deforms the AdS-dS-wormhole only in the far future, removing the regulator must send the total entropy surfaces to the future conformal boundary.  And since their area remains bounded, they cannot approach the interior of $\mathscr{I}_\mathrm{dS}$.  But there is no need to regulate the spacetime far from $\mathscr{I}_\mathrm{dS}$, so any limiting surface can have no finite separation from $\mathscr{I}_\mathrm{dS}$.  The limiting total entropy surface must thus lie at one of the edges $\E$ in figure \ref{fig:wormhole} that mark the boundary between $\mathscr{I}_\mathrm{dS}$ and the singular part of the future conformal boundary.   For similar reasons we expect that studying minimal surfaces on achronal surfaces $\Sigma$ converging to the future conformal boundary in the unregulated AdS-dS-wormhole  will also lead to effective maximin surfaces located at one of the edges~$\E$; i.e., that the maximin procedure naturally defines a surface in the conformal extension $\overline M$.  In examples with right/left symmetry we should assign two surfaces, one at each edge.  In other cases the choice of left edge vs. right\footnote{Here we assume that $\mathscr{I}_\mathrm{dS}$ is connected.} will depend on details of the AdS-dS-wormhole, though we expect that it will not depend on the choice of regulator.

\subsection{Implications for entanglement}
\label{subsec:implications}

Let us now return to the discussion of entanglement.  We begin with toroidal AdS-dS-wormholes in which the translation symmetry is compact.  We argued above that the corresponding dual CFTs have non-zero leading-order mutual information.  We also showed in section \ref{sec:theorem} that $I(A,B) \approx 0$ for regions $A$,~$B$ on opposite boundaries having sizes much smaller than the size of the torus.  However, the mutual information can be non-zero at this order for $A,B$ sufficiently large.

To discuss the uncompactified case we take the large-torus limit while holding fixed the size of our regions $A$,~$B$.  The part of the opposite CFT strongly entangled with $A$ then recedes to infinity, while the total mutual information per unit area between the two CFTs remains constant.  This suggests that one think of each infinite plane in the non-compact case as the limit of entire tori so that, although finite-sized subregions in opposite CFTs have no leading-order entanglement, the resulting planar CFTs retain finite leading-order mutual information per unit area; i.e., although correlations recede to infinity we do not allow any information to be lost in taking the limit\footnote{A theorem of \cite{Araki:1976zv} (Lemma 3, Remark 1) shows that one may successfully approximate any relative entropy defined on a von Neumann algebra by describing this algebra as a limit of smaller algebras.  The same thus holds for mutual information.  The above interpretation is consistent with this theorem, as the algebra it assigns to the plane effectively contains many operators ``at infinity'' which are not limits of operators in finite regions.  This ``algebra at infinity'' corresponds to the distant parts of the finite tori used to take the limit.}.  Repeating this discussion for effective total entropy surfaces lying in the future conformal boundary of the toroidal wormholes leads us to consider similar effective total entropy surfaces for the planar wormholes, and of course the limit of empty sets remains empty.  For the finite tori, the former compute the entropy of each boundary separately while the latter (empty set) surfaces compute the total entropy of both boundaries together.  So it is in the above sense that, in the case of non-compact cross-section, our effective HHRT surface and the empty set respectively compute the leading order  total entropy of each CFT separately and for the joint state on the pair of CFTs. 

While the above notion of limit is essentially unique for finite-sized regions $A$ and $B$, it should be mentioned that there is an alternate way of interpreting what is meant by the limiting planar CFTs taken as wholes.  In this second interpretation, each entire plane is the limit of a family of additional (larger) subregions of the growing tori. These larger regions are taken to grow in size without bound, but at a rate much slower than the size of the torus itself.  In other words, one ``zooms in'' on a smaller and smaller fraction of the torus as the torus grows.  Since each resulting plane is built from the limit of ``small'' regions of the large-but-finite tori, the total leading-order mutual information between the two CFTs must vanish.   For a finite torus, the corresponding HHRT surfaces are then anchored to ``small'' regions of the boundary $\mathscr{I}_\mathrm{AdS}$ and cannot enter the past of $\mathscr{I}_\mathrm{dS}$.  Moreover, two such regions on opposite boundaries are not homologous and require distinct HHRT surfaces.  Taking the large torus limit then implies that we continue to assign the total entropy of each planar CFT a distinct HHRT surface\footnote{This may be only an effective surface in a sense similar to that of the \HHRT~proposal.} lying entirely on its side of the wormhole, and that the union of these surfaces describes the total entropy of the two CFTs together.  The total leading-order mutual information between the two CFTs then vanishes as desired under the alternate interpretation just described for the limiting planar CFTs.

The highly delocalized entanglement characteristic of toroidal AdS-dS-wormholes thus leads to two physically-distinct notions of the planar limit, both described by the same limiting (planar AdS-dS-wormhole) spacetime.  The preceding analysis suggests that non-compact wormholes generally admit at least two correspondingly distinct interpretations of the homology constraint, associated with different possible roles being played by the region ``at infinity'' in directions transverse to the dimensions displayed in our figures.   But we leave further development of this proposal for future work and content ourselves here with the discussions above.

\section{Complex wormhole-spanning Surfaces?}
\label{sec:complexHM}

While we see no inherent inconsistencies in the CFT entropies predicted by \HHRT,  the infinite area of real wormhole-spanning $\overline{\mathrm{HHRT}}$ surfaces makes our AdS-dS-wormholes a natural context in which to investigate further possible improvements.  In particular, one might ask if complex extremal surfaces could play a role.  This is suggested by the superficial analogy with the geodesic approximation to two-point functions where a lack of real geodesics does indeed indicate the importance of complex ones \cite{Fidkowski:2003nf}; see also \cite{Kraus:2002iv,Festuccia:2005pi,Balasubramanian:2012tu}  for more general discussions.  It would be very interesting to investigate complex extremal surfaces in particular example AdS-dS-wormholes (as done for static planar black holes in \cite{Fischetti:2014zja}) and to see if the results inform any of the conceptual puzzles associated with the use of complex surfaces (see again \cite{Fischetti:2014zja} for discussion).  However, since the cut-and-paste spacetimes of section \ref{sec:cutpaste} are not analytic, it is unclear in what complexification such complex extremal surfaces might live.  Indeed, analogy with the geodesic approximation to two-point functions raises the question of whether any HHRT-like prescription can apply to geometries that are not analytic; see e.g. \cite{Louko:2000tp}.

We thus save analysis of complex surfaces in actual AdS-dS-wormholes for future work and make no attempt to study them here. Instead, we briefly discuss complex codimension-2 surfaces in pure de Sitter space.  This section thus represents a slight aside from the main theme of this work and may be skipped without loss of continuity.  In pure de Sitter settings analogous to spanning our wormholes, real such surfaces again do not exist.  But we shall see that complex surfaces are readily found.

Of course, the existence of complex such surfaces does not immediately imply their relevance to the computations at hand.  For example, if they describe complex saddles approximating some path integral, complex surfaces will contribute only if one can appropriately deform the contour of integration to include them.  While it is unclear how to analyze this in detail for the entropy problem, it is interesting to consider the superficially-related problem of computing free-field two-point functions in the Bunch-Davies vacuum using the geodesic approximation. Using an expansion of this two-point function from appendix \ref{app:2pt}, we show explicitly below how it is given by an infinite set of complex geodesics in~dS$_3$ and that these geodesics lie on an infinite number of sheets of the associated Riemann surface in the sense of \cite{Fischetti:2014zja}.

\subsection{Complex Extremal Surfaces in dS}

\begin{figure}[t]
\centering
%
%
%
%
%
\includegraphics[page=11]{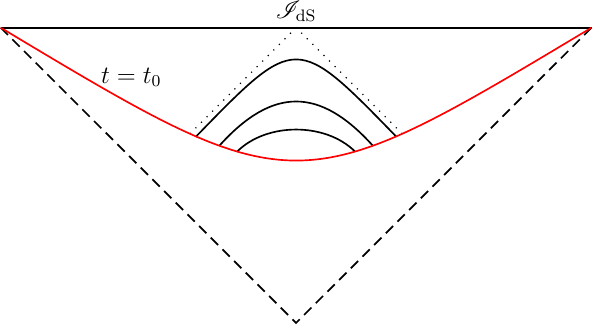}
\caption{The cosmological patch of de Sitter spacetime.  The solid black curves are sketches of extremal surfaces ending on a~$t = t_0 = \mathrm{const}$ slice, shown as a solid red curve.  As the endpoints of the surfaces are taken farther apart, the surfaces approach the dotted null curve.  At even larger separations real extremal surfaces cease to exist.}
\label{fig:deSitter}
\end{figure}

It is well known that pure de Sitter space contains pairs of points that cannot be connected by geodesics (see e.g. \cite{Hawking:1973uf}).  Indeed, geodesics tend to bend down and away from future infinity, as shown in figure~\ref{fig:deSitter}.  So if the ends of the extremal surface are taken far enough apart, the geodesic becomes null and ``bounces'' off future infinity in a manner pictorially similar to the bouncing geodesics of AdS-Schwarzschild \cite{Fidkowski:2003nf,Fischetti:2014zja} -- though the null limit of bouncing geodesics retains finite length in~dS$_{d+1}$ while it vanishes in AdS-Schwarzschild as measured from any finite points in the spacetime.  Real geodesics cease to exist when the separation is increased beyond this critical point, leaving only complex ones.  This occurs in particular for $d=2$, where geodesics are codimension-2 extremal surfaces.  Extremal surfaces of any codimension turn out to behave similarly for all $d$, though the area diverges in the null limit for extremal surfaces whose dimension exceeds 1 (i.e., for any case except geodesics).

We now study this phenomenon in detail for a class of codimension-2 extremal surfaces analogous to the would-be wormhole-spanning surfaces of section \ref{sec:theorem}.  Below we anchor our surfaces at the de Sitter horizon as opposed to at a spacetime boundary.  This allows us to work entirely in the dS patch.  We study pure de Sitter for simplicity, but analogous results should also hold for patch II as defined in section \ref{sec:cutpaste}.

Consider the inflating spatially-flat patch of pure de Sitter in the familiar coordinates where the metric takes the form
 \be
\label{eq:deSitterflat}
ds^2 = -dt^2 + e^{2Ht} d\vec{x}_d^2 \equiv -dt^2 + e^{2Ht} \left(dr^2 + (dx^1)^2 + (d\vec{x}^\parallel_{d-2})^2\right).
\ee
We take our surfaces to be anchored on infinite strips defined by~$t = t_0$,~$r = \pm L/2$,~$x^1 = \mathrm{const.}$ Conservation of $x^1$ momentum implies that~$x^1$ remains constant across the entire extremal surface.  The surface can thus be parametrized by the coordinates~$x^\parallel$ and a yet-to-be-specified parameter~$\lambda$; i.e.,~$(t,r,x^1,x^\parallel) = (t(\lambda), r(\lambda), \mathrm{const.}, x^\parallel)$.  The resulting area functional is
\be
\label{eq:area}
A = V_{d-2} \int d\lambda \, e^{(d-2)H t} \sqrt{e^{2Ht} \dot{r}^2 - \dot{t}^2} \equiv V_{d-2} \int d\lambda \, \mathcal{L}\left(t,r,\dot{t},\dot{r}\right),
\ee
where~$V_{d-2} \equiv \int d^{d-2} x^\parallel$ is the volume of the space spanned by the~$x^\parallel$ coordinates.

\begin{figure}[t]
\centering
\subfigure[]{
\includegraphics[width=0.4\textwidth]{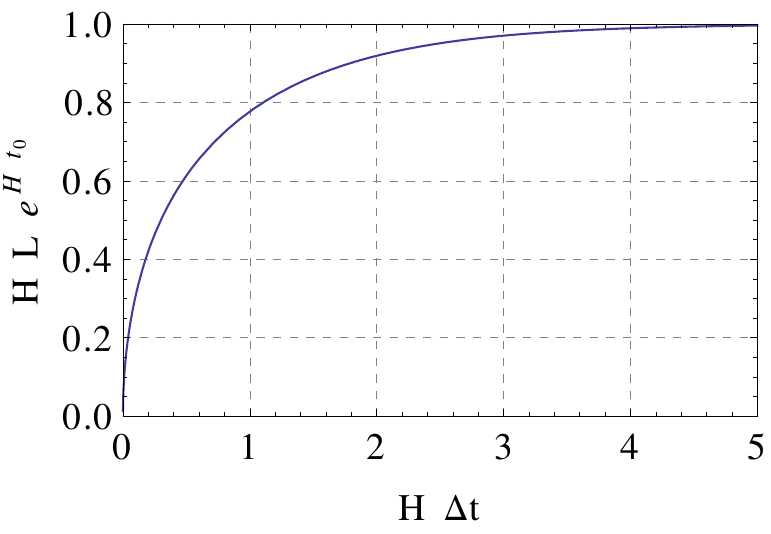}
\label{subfig:Lvst}
}
\quad
\subfigure[]{
\includegraphics[width=0.4\textwidth]{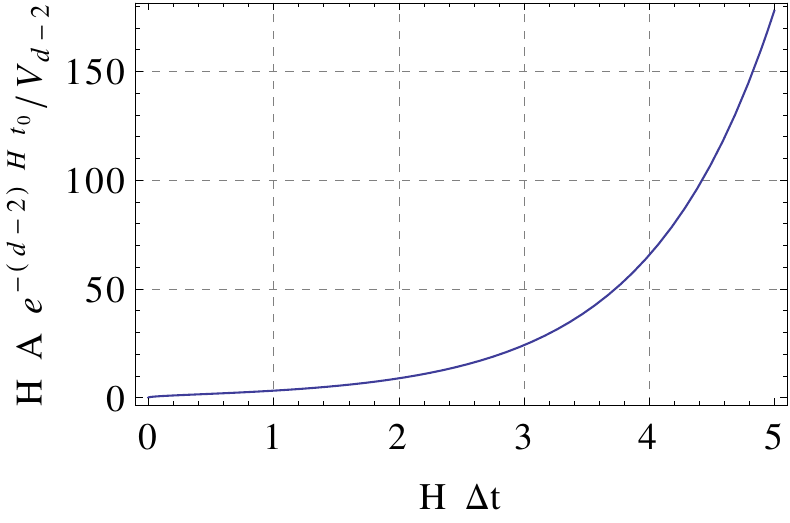}
\label{subfig:Avst}
}
\caption{The coordinate displacement~$L$ (a, at left) and the area~$A$ of codimension-2 extremal surfaces (b, at right) in pure dS as functions of~$\Delta t$.  The plots show results for~$d = 3$, though the qualitative behavior is unchanged for~$d \geq 3$ (for~$d = 2$, the area remains finite at large~$H \Delta t$).  Note that~$L$ approaches a constant~$e^{-H t_0}/H$ at large~$H \Delta t$, consistent with the fact that real extremal surfaces do not exist for larger~$L$.}
\label{fig:realLA}
\end{figure}

Since the effective Lagrangian~$\mathcal{L} = e^{(d-2)H t} \sqrt{e^{2Ht} \dot{r}^2 - \dot{t}^2}$ contains no explicit dependence on~$r(\lambda)$, there is a conserved conjugate momentum
\be
\label{eq:momentum}
p = \frac{\partial \mathcal{L}}{\partial r} = \frac{e^{dHt} \dot{r}}{\sqrt{e^{2Ht} \dot{r}^2 - \dot{t}^2}}.
\ee
Choosing the parameter~$\lambda$ so that~$e^{2Ht} \dot{r}^2 - \dot{t}^2 = 1$, we obtain
\begin{subequations}
\label{eqs:rdottdot}
\bea
\label{subeq:rdot}
\dot{r} &= e^{-Ht_*} e^{-dH(t-t_*)}, \\
\label{subeq:tdot}
\dot{t}^2 + V_\mathrm{eff}(t) &= 0,
\eea
\end{subequations}
in terms of an effective Newtonian potential
\be
V_\mathrm{eff}(t) = 1 - e^{-2(d-1) H(t-t_*)}.
\ee
Here~$t_* \equiv \ln p/((d-1)H)$ is the real root of~$V_\mathrm{eff}(t)$ and describes the turning point of real extremal surfaces.  Relating $t_*$ to the coordinate displacement~$L$ between the anchor points through
\be
2L = 2\int_{t_0}^{t_*} \frac{\dot{r}}{\dot{t}} \, dt
\ee
yields
\begin{multline}
\label{eq:dSL}
e^{H t_0} L = \frac{i}{d H} \, e^{-H \Delta t} \left[e^{d H \Delta t} \, {}_2 F_1\left(\frac{1}{2}, \frac{d}{2(d-1)}; \frac{3d-2}{2(d-1)}; e^{2(d-1)H \Delta t}\right) \right. \\ \left.- \, {}_2 F_1\left(\frac{1}{2}, \frac{d}{2(d-1)}; \frac{3d-2}{2(d-1)}; 1\right) \right],
\end{multline}
where~$\Delta t \equiv t_* - t_0$ and ${}_2 F_1$ is the ordinary hypergeometric function written using standard conventions (e.g.~\cite{Prudnikov:1986}).  Likewise, the area~\eqref{eq:area} becomes
\begin{multline}
\label{eq:dSarea}
e^{-(d-2)H t_0}A = \frac{2iV_{d-2}}{(d-2) H} \left[e^{(d-2)H \Delta t} \, {}_2 F_1\left(\frac{1}{2}, -\frac{d-2}{2(d-1)}; \frac{d}{2(d-1)}; 1\right) \right. \\ \left.- \, {}_2 F_1\left(\frac{1}{2}, -\frac{d-2}{2(d-1)}; \frac{d}{2(d-1)}; e^{2(d-1) H \Delta t} \right) \right].
\end{multline}
These~$L$ and~$A$ are plotted in figure~\ref{fig:realLA} as functions of~$\Delta t$.  It is clear that $L$ approaches a finite value as~$\Delta t \to \infty$.  Indeed, expanding~\eqref{eq:dSL} at large~$\Delta t$, we obtain
\be
e^{H t_0} L = \frac{1}{H} + \mathcal{O}\left(e^{-H \Delta t}\right).
\ee

As advertised, the surface becomes null in the limit $t_* \rightarrow \infty$ ($L \rightarrow H^{-1}e^{-Ht_0})$.  As shown in figure \ref{fig:deSitter}, for $L < H^{-1}e^{-Ht_0}$ the entire extremal surface lies within the past light cone of a set on $\mathscr{I}_\mathrm{dS}$ of vanishing length in the $r$-direction.  This is in fact required by the same reasoning as in section \ref{sec:theorem}.  Such arguments imply that a null surface fired orthogonally from an extremal surface can intersect $\mathscr{I}_\mathrm{dS}$ only in some zero-measure set.  But continuity requires that the image of our null geodesics on $\mathscr{I}_\mathrm{dS}$ must span some interval in $r$.  Thus the length of this interval must vanish.

\begin{figure}[t]
\centering
\includegraphics[width=0.35\textwidth]{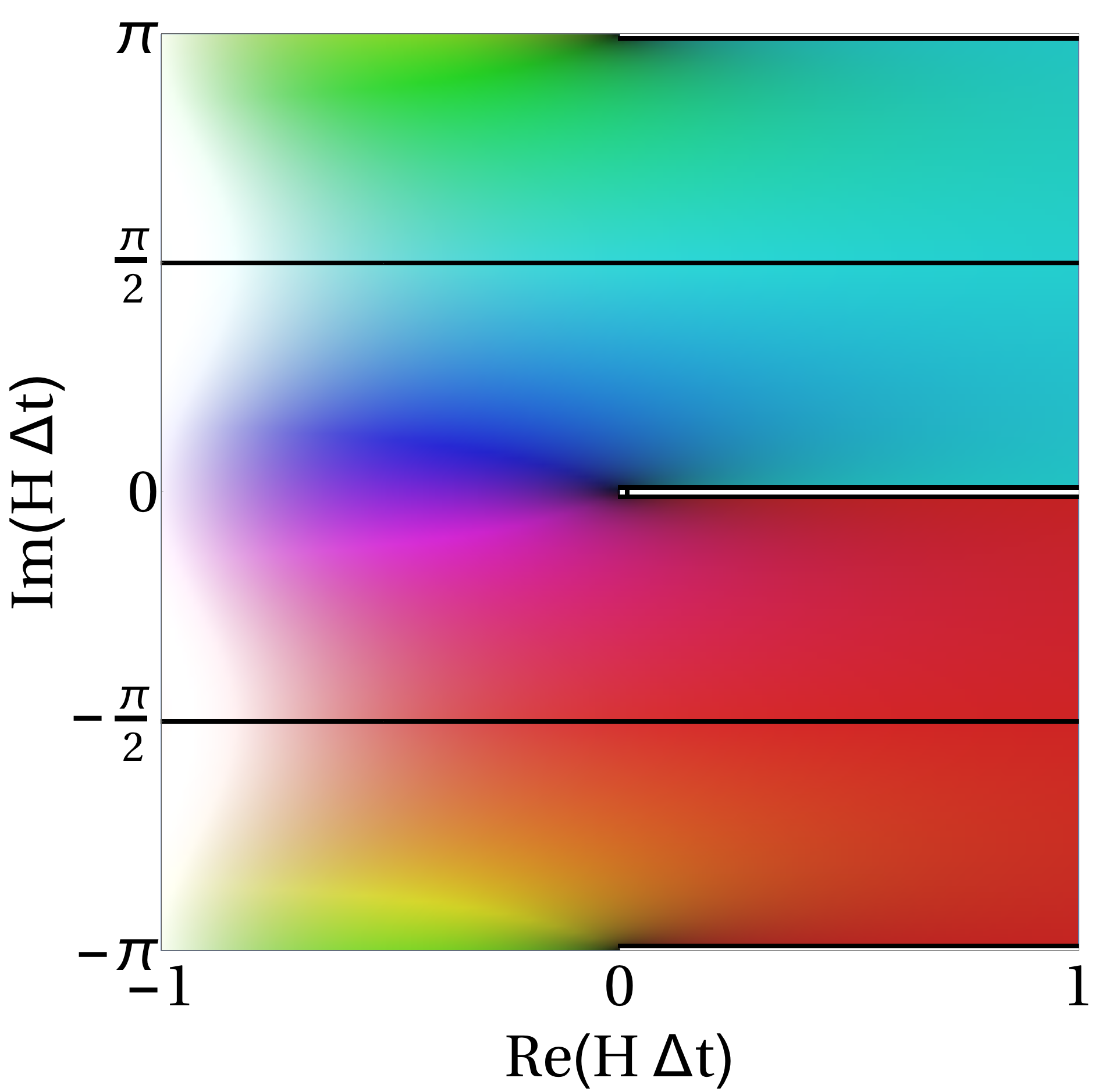} \quad
\includegraphics[width=0.35\textwidth]{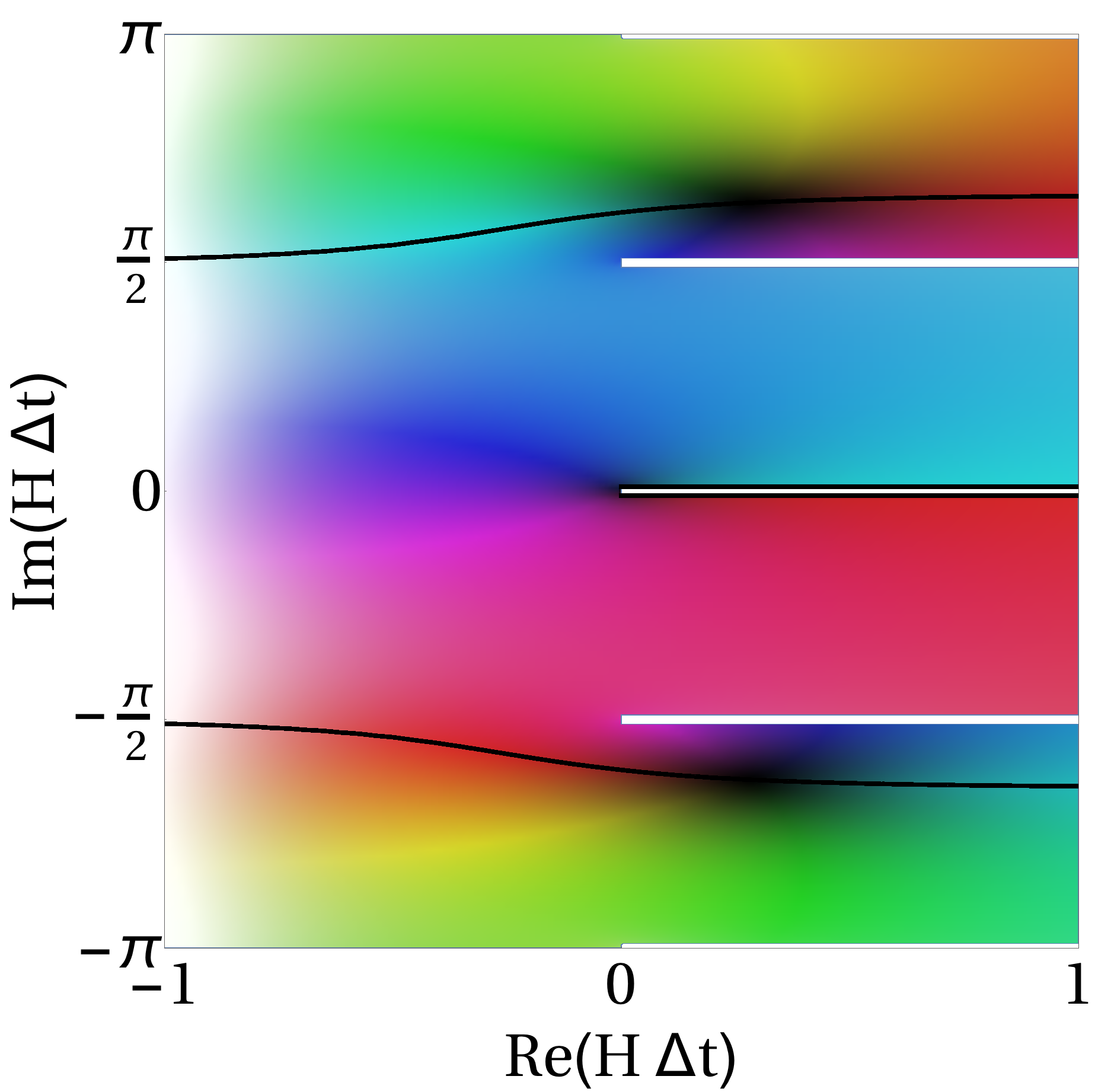} \\
\includegraphics[width=0.35\textwidth]{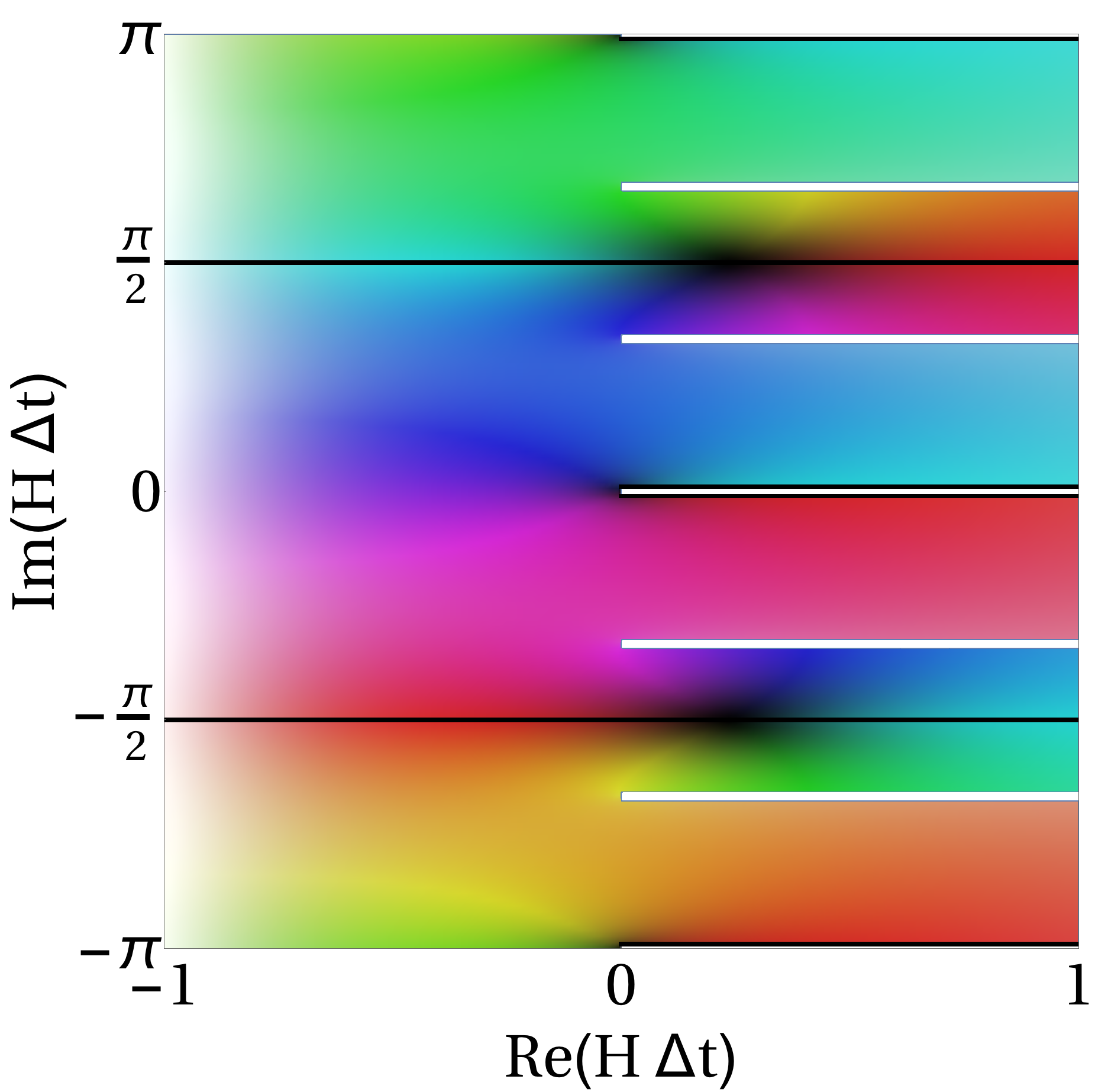} \quad
\includegraphics[width=0.35\textwidth]{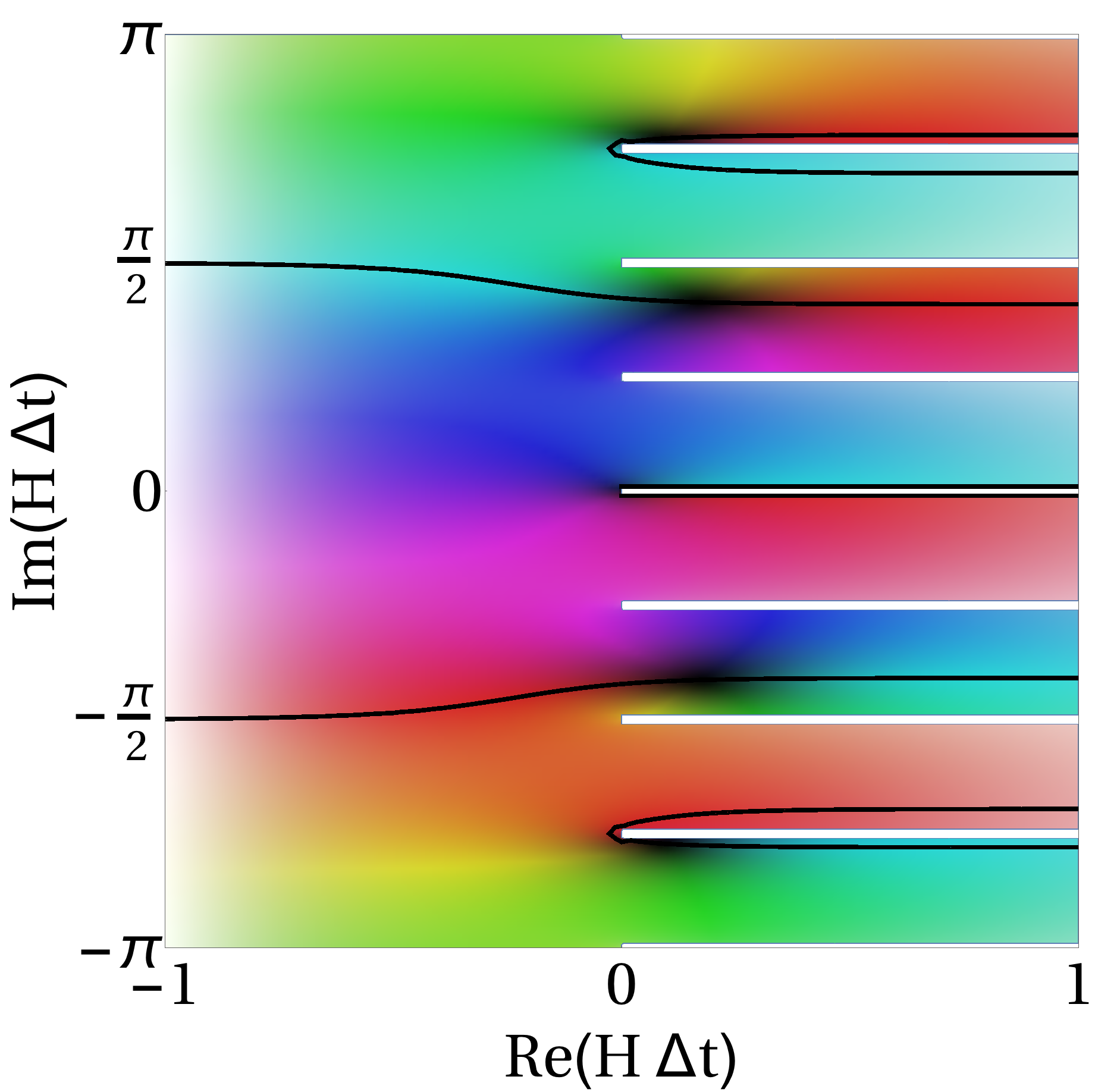} \\
\includegraphics[width=0.35\textwidth]{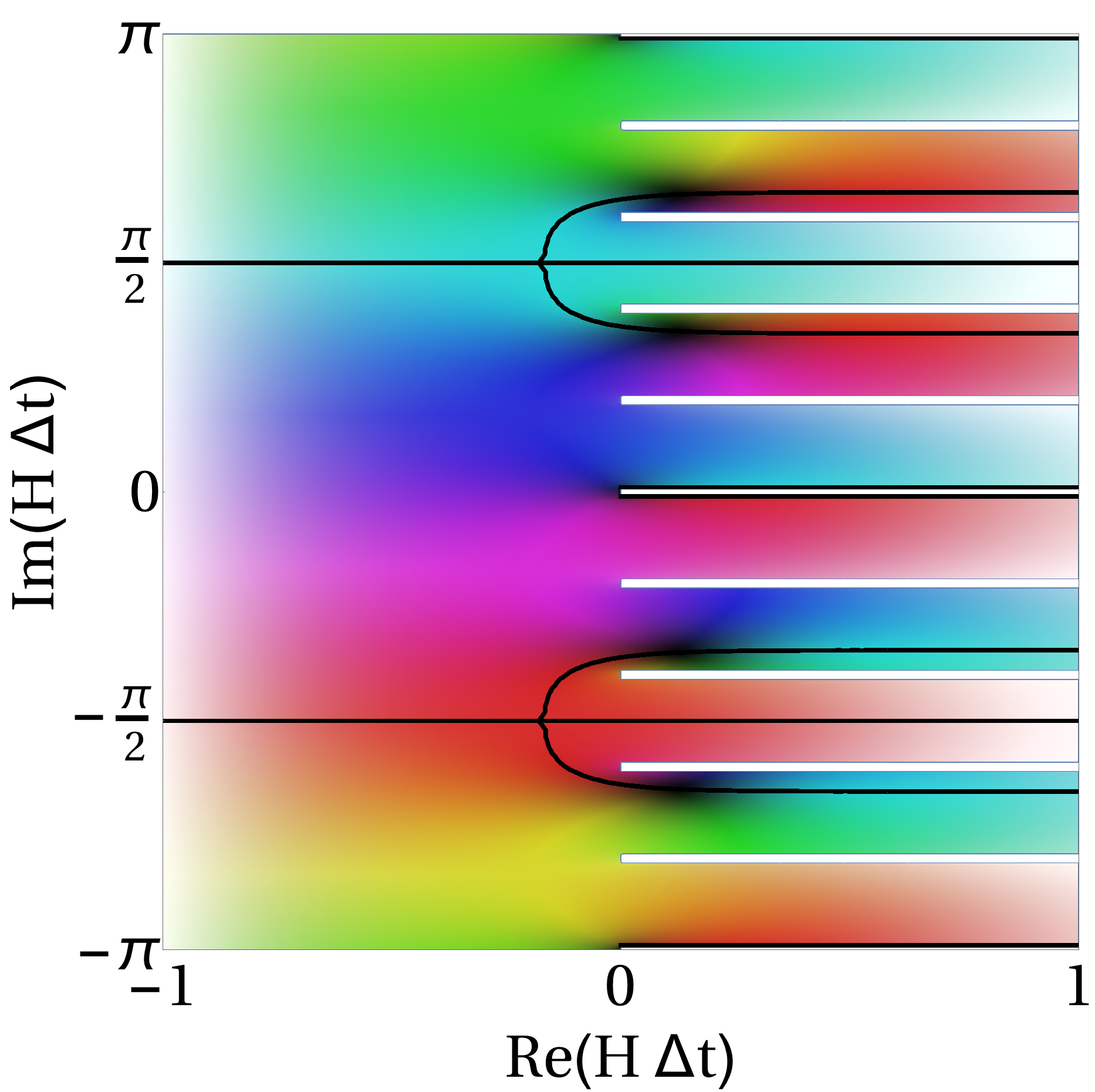} \quad
\includegraphics[width=0.35\textwidth]{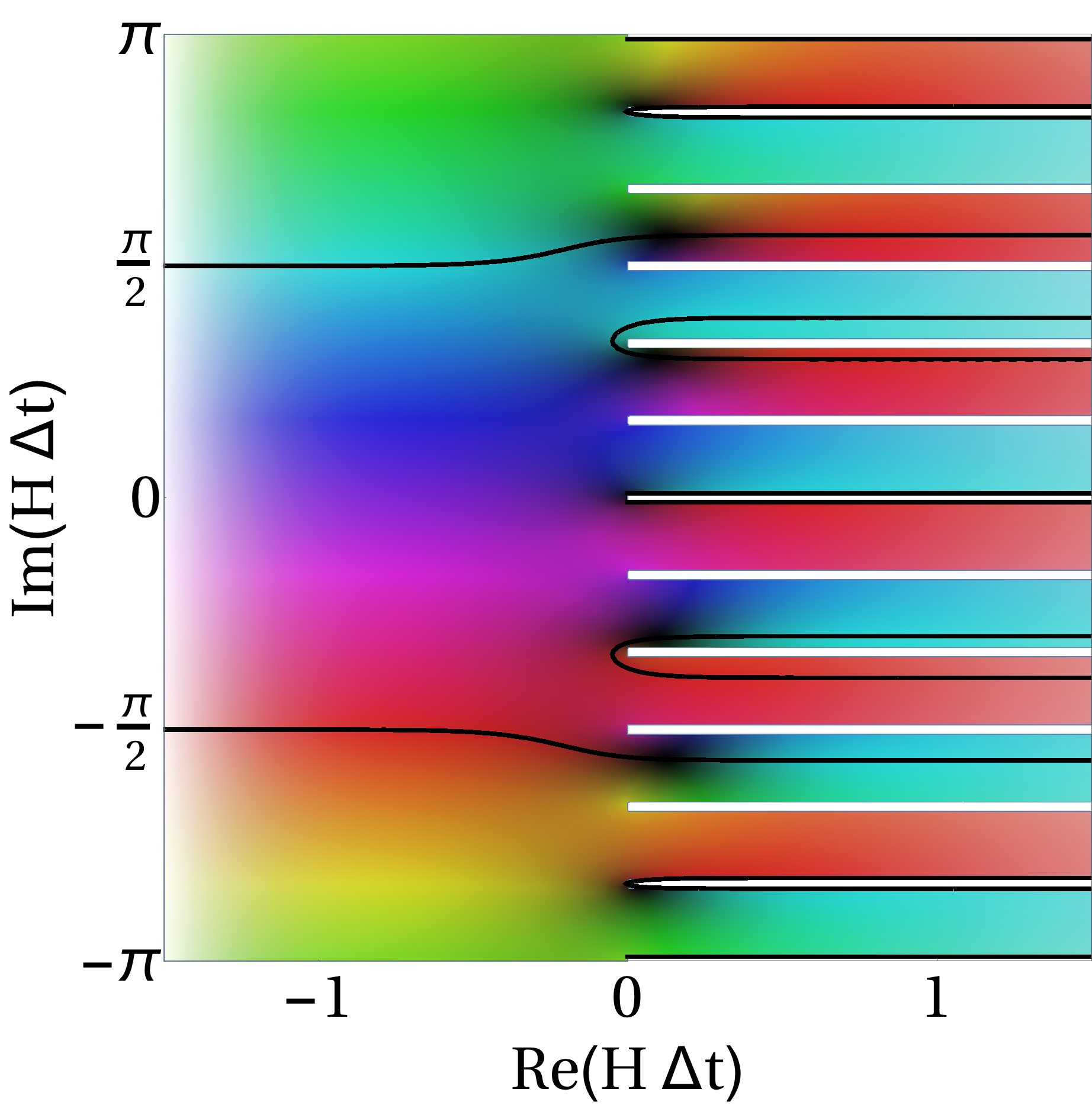}
\caption{$L$ as a function of complex~$\Delta t$ in~dS$_d$.  From left to right and top to bottom, the plots show~$d = 2, 3, 4, 5, 6, 7$.    Hue indicates~$\arg(L)$ (with real positive [negative]~$L$ in red [turquoise]), while shade indicates the magnitude of~$|L|$ (with~$|L| = 0$ in black and increasing~$|L|$ in lighter shades).  The white horizontal strips mark the locations of branch cuts, and the black lines are contours along which~$L$ is real.  $|L|$ is bounded in the right half-plane, but grows without bound in the left half-plane; thus the only contours that can reach arbitrarily large real~$L$ are the two complex ones that run to large negative~$\Re(\Delta t)$.}
\label{fig:complexL}
\end{figure}

In contrast, for $L > H^{-1}e^{-Ht_0}$ causality would require this interval have non-vanishing length.  So real codimension-2 surfaces can no longer exist. But it is straightforward to find complex extremal surfaces in this regime (and indeed for arbitrary~$L$ when $d > 2$).  One simply analytically continues expressions~\eqref{eq:dSL} and~\eqref{eq:dSarea} to the entire complex~$\Delta t$-plane.  From~\eqref{eq:dSL} we see that~$L$ is periodic in~$\Delta t$ with period~$2\pi i/H$, so it suffices to study~$L$ in a finite strip around the real axis.  Figure~\ref{fig:complexL} shows the complex-valued function~$L(\Delta t)$ in this strip for~$d=2,3,4,5,6,7$; in particular, we indicate contours along which~$L$ is real.  One of these runs along the real positive~$\Delta t$-axis, looping tightly around the branch cut, but the others lie at complex $\Delta t$. We see that one may obtain large positive $L$-values by taking $\rm{Re} \ \Delta t$ large and negative along one of the real $L$ contours in the lower half plane.  For $d > 2$ this contour clearly also reaches $L=0$ (and in fact passes to negative $L$), providing a complex extremal surface for all physically relevant $L$.

In parallel with the results of~\cite{Fischetti:2014zja} for black holes, we expect additional contours of real~$L$ to exist on other sheets of the Riemann surface for~$L(\Delta t)$. This function and its Riemann surface is defined by analytic continuation through the branch cuts in figure~\ref{fig:complexL}. The branch points are of logarithmic type for $d > 2$ where they lead to an infinite number of sheets.  The $d=2$ case is special in that the branch points of $L(\Delta t)$ are only two-sheeted square-root type branches; though in that case there are additional infinite-sheeted logarithmic branchings of the physically-interesting function $A(L)$ that make the overall structure much the same.

Of course, the mere existence of complex extremal surfaces need not imply that they are relevant to our study of entropy. For bulk spacetimes constructed via some Euclidean path integral, one could plausibly use analytic continuation and the argument of \cite{Lewkowycz:2013nqa} to write the desired entropy in terms of extremal surfaces.  But for a given complex extremal surface to appear in this calculation it must be possible to appropriately deform the original contour of integration.  A priori, this is far from guaranteed -- though since there are no real extremal surfaces for $L > H^{-1}e^{-Ht_0}$, any contours that are allowed must be complex.

For $d=2$ our codimension-2 surfaces are geodesics and the area becomes a length.  As noted earlier, the length of bulk geodesics can also be used to approximate two-point functions of CFT operators with large dimension (so long as it is still small enough to ignore gravitational back-reaction).  This is of course closely related to our entropy problem, since entropy can be calculated from the two-point function of twist operators \cite{Holzhey:1994we}.  These twist operators do indeed have large dimension -- though, since acting with appropriate twist operators is equivalent to replicating the entire large $N$ CFT in the sense of the replica trick,  their dimension is in fact large enough the gravitational back-reaction is generally non-trivial.  So while the two calculations are not precisely the same, it is interesting to write the well-known exact two-point functions in dS${}_3$ as a sum over complex geodesics.  This result is presented in appendix~\ref{app:2pt}, which finds this sum to use an infinite number of terms from an infinite number of sheets of the Riemann surface for $A(L)$.  The analogue for $d >2$ would be to use an infinite number of complex geodesics on an infinite number of sheets of the Riemann surface for $L(\Delta t)$.  So at least in this context there is no problem deforming the relevant path integral to take advantage of complex saddles.  It is tempting to suggest that related contours will be relevant for studying the entropy of AdS-dS-wormholes, leading to non-zero leading-order mutual information between localized regions on opposite boundaries.

\section{Discussion}
\label{sec:discussion}

This work considered two-sided AdS-dS-wormholes, which are spacetimes that contain a region of unbounded inflation.  In particular, the future conformal boundary of the wormhole interior contains a smooth spacelike piece $\mathscr{I}_\mathrm{dS}$ as shown in figure \ref{fig:wormhole}.  Explicit examples satisfying the null energy condition were constructed in section \ref{sec:cutpaste}.  While our smooth examples contain Cauchy horizons, we expect such solutions to be unstable to decay into a more generic class of AdS-dS-wormholes which otherwise retain all of the properties discussed below.

Our main result is that these geometries fail to admit HHRT surfaces (i.e., possibly non-minimal Hubeny-Rangamani-Takayanagi surfaces with the homology constraint emphasized by Headrick) that would exist in more familiar black hole spacetimes.
In particular, section \ref{sec:theorem} showed that no HHRT surface can span the wormhole, connecting one side to the other.  Instead, HHRT surfaces for the associated entropy problems must be disconnected, with one piece on each side of the wormhole.  Section \ref{sec:total} showed that certain of our wormholes have plane-symmetric HHRT surface homologous to an entire boundary -- which we termed total entropy surfaces -- there is also a large class that do not.  We suggested that less-symmetric such surfaces also fail to exist, so that there are no extremal codimension-2 surfaces in the entire homology class.  If so, the HHRT proposal becomes ill-defined and requires improvement.  We also gave related examples where HHRT surfaces would exist but give physically incorrect results.

The conceptually-simplest possible changes to HHRT were discussed in section \ref{sec:total}.  These involve first regulating the dS-wormhole by allowing only a finite amount of inflation.  After the inflating phase, the wormhole is required to collapse to a future singularity; see appendix \ref{app:regulated} for details.  We argued that, at least in our examples, the limit in which the regulator is removed gives natural wormhole-spanning and total entropy HHRT surfaces lying in the future conformal boundary; i.e., they lie in the conformally extended spacetime $\overline M$ instead of the physical spacetime $M$.  In the wormhole-spanning case this surface has infinite area and so is never the minimal-area surface.  But in the total entropy context any limiting surface must have finite entropy density consistent with the CFT density of states.  This regulate-and-take-limits approach was called $\overline{\rm{HHRT}}$.  But we did not investigate the convergence of these limits in detail, so it remains to determine the extent to which they are well-defined.

We also suggested an extended maximim prescription $\overline{\rm maximin}$ that takes limits directly in the unregulated wormhole spacetime and may give results identical to $\overline{\rm HHRT}$.  The $\overline{\rm maximin}$ procedure clearly assigns well-defined (though perhaps infinite) area to each entropy problem, and in appropriate cases may also yield a well-defined $\overline{\rm maximin}$ surface in $\overline M$.  But we did not analyze precisely when this surface construction succeeds, and it again remains to study when this area will agree with regulate-and-take-limits procedures.

Under either $\overline{\rm HHRT}$ or $\overline{\rm maximin}$ one finds that toroidal AdS-dS-wormholes are dual to highly entangled pure states on a pair of CFTs, and that this remains true for planar CFTs obtained through an appropriate large-torus limit\footnote{Though there is another large-torus limit where it does not.  Both limits are described by the same planar AdS-dS-wormhole but with different notions of the homology constraint.  See section \ref{subsec:implications}.}.  But the associated mutual information is as delocalized as possible.  In particular, for CFTs on infinite spacetimes, the leading-order mutual information vanishes between any finite-sized regions $A,B$ of opposite CFTs.

A strictly vanishing mutual information between finite-sized subregions would contradict the non-vanishing correlators $\langle {\cal O}_{CFT_1} (x_1,t_1) {\cal O}_{CFT_2} (x_2,t_2) \rangle$ associated with taking appropriate boundary limits of bulk two-point functions\footnote{\label{note:bulk2pt}If our wormhole can be found as the Wick rotation of a saddle that dominates the Euclidean path integral, this integral defines a state in which the correlator can be computed using the geodesic approximation (and where it will be non-zero). But in any case the linearized bulk equation of motion would allow the above CFT correlator to vanish identically only if the corresponding bulk correlator $\langle \phi (x_1,r_1,t_1) \phi (x_2,r_2,t_2) \rangle$ vanishes for all $(x_1,r_1,t_1)$ in the left region I and all $(x_2,r_2,t_2)$ in the right region I.  This is a very fine-tuned property and we are free to consider bulk quantum states on our wormhole background for which it does not hold.}; see e.g. \cite{2008PhRvL.100g0502W}.  But the claim is only that the mutual information vanishes at leading order in large $N$, so some finite mutual information may remain.  Indeed, according to \cite{Faulkner:2013ana} (see also \cite{Engelhardt:2014gca}) it is precisely the $O(1)$ correction that is encoded in the state of bulk quantum fields to which the supergravity approximation applies.  Such fields are dual to CFT operators whose dimensions are not too large.  The implication is thus only that generic operators of large dimension (e.g., of order $N^2$ in 3+1 ${\cal N}=4$ super Yang-Mills) have vanishing correlators between the two CFTs.

We see no inherent contradiction with this interpretation. Indeed, the physics is quite similar to that naively obtained from the extreme limit of Reissner-Nordstr\"om black holes.  There the large area of wormhole-spanning surfaces is associated with the infinite throat that develops at zero temperature ($T=0$). The most apparent difference is that for Reissner-Nordstr\"om the two-boundary spacetime becomes disconnected at $T=0$, making it somewhat more natural to consider quantum states of the linearized bulk fields having vanishing correlators between the two sides.  But there are also states with non-vanishing correlators, and for fields with fine-tuned values of the bulk charge and mass such states are in fact naturally constructed by the bulk path integral dual to charged thermofield-double states in the CFT \cite{Andrade:2013rra}.  A more critical difference may be that small $T$ Reissner-N\"ordstrom black holes tend to be unstable in top-down models, while causality forbids any instability of our exterior (the left and right copies of region I) being activated by starting inflation in the interior of our wormhole.

Intriguingly, the physics is also quite similar to that expected for generic entangled states (see e.g \cite{Hartman:2013qma,Marolf:2013dba,Shenker:2013yza,Leichenauer:2014nxa} for holographic discussions).  This is even more so when one chooses quantum states for the bulk fields where $\langle {\cal O}_{CFT_1} (x_1,t_1) {\cal O}_{CFT_2} (x_2,t_2) \rangle =0$ (see again footnote \ref{note:bulk2pt}).    The one point of tension is that \cite{Shenker:2013yza} predicted wormholes associated with such generic states to have time-independent interiors -- though there is no actual contradiction so long as all implications for the CFTs remain time-independent.

It is possible that such physical predictions are correct and will provide insights into the holographic description of inflation.  But the paucity of real codimension-2 surfaces makes our AdS-dS-wormholes a natural context in which to investigate further possible modifications of HHRT.    For example, one might ask if our wormholes might have no dual interpretation at all, or more conservatively if dual descriptions might require more than just a pair of CFTs; e.g., despite the HHRT claim that the state on both CFTs is pure at leading order in large $N$, one might suppose that the natural two CFTs are both highly entangled with some third system. This latter option would be analogous to the mixed-state proposal of \cite{Freivogel:2005qh}, and the third system might correspond to the superselection sectors of \cite{Marolf:2008tx,Marolf:2012xe}.  The present understanding of gauge/gravity duality is sufficiently coarse that we cannot exclude such suggestions, though as in \cite{Freivogel:2005qh}, it is natural to take the constructions of \cite{Kachru:2003aw} and related work as suggesting that a dual interpretation does in fact exist. And if one can construct our wormholes from (due to the lack of time-symmetry, complex) saddle points for Euclidean path integrals then one should be able to argue as in the thermofield-double discussion of \cite{Maldacena:2001kr} that it is given by a pure state on two CFTs.  Indeed, one should then also be able to argue as in \cite{Lewkowycz:2013nqa} that something like HHRT does in fact hold.

The discussion of complex saddles naturally motivates a milder possible modification of HHRT that, at least in analytic spacetimes, would make use of complex extremal surfaces in addition to real ones.   For bulk black holes dual to thermofield double states this option was studied in \cite{Fischetti:2014zja}, and for AdS-dS-wormholes it was briefly addressed in section \ref{sec:complexHM}. In particular, noting that HHRT is superficially similar to the the geodesic approximation for two-point functions motivated a study of this latter context.  We considered the case of dS${}_3$ -- where geodesics are also codimension-2 extremal surfaces -- and found complex geodesics to be critical in constructing a stationary-phase approximation to the exact result.  In particular, in the two-point function calculation it appears that one can deform the contour of integration to take advantage of complex geodesics living on an infinite number of sheets of the associated Riemann surface.

It would be very useful to study complex HHRT surfaces in full AdS-dS-wormholes.  One would specifically like to understand whether the results might shed light on the confusions surrounding the use of complex surfaces that were discussed in \cite{Fischetti:2014zja}.  Unfortunately, since the cut-and-paste examples of section \ref{sec:cutpaste} are not analytic, the complexification of these particular spacetimes is far from unique and any notion of complex surfaces may be ill-defined.  This places a detailed analysis of complex extremal surfaces in any AdS-dS-wormhole beyond the scope of this work, making it an interesting challenge for future investigation.

\section*{Acknowledgements}

The authors wish to thank Ahmed Almheiri and Mukund Rangamani for useful discussion and comments.  This project was supported in part by the National Science Foundation under Grant No PHY11-25915, by FQXi grant FRP3-1338, and by funds from the University of California.

\appendix

\section{Shell Stress Tensors}
\label{app:shells}

We now compute the stress tensors on the null shells of section \ref{sec:cutpaste}. Following~\cite{Poisson}, we embed each shell in each associated patch of the spacetime via parametric relation~$x^\alpha = X^\alpha(y^a)$, where the~$y^a$ are a set of~$d$ coordinates on the shell and the~$x^\alpha$ are the spacetime coordinates of the patch in which the shell is to be embedded.  We take $d-1$ of the~$y^a$ to be the transverse coordinates~$x^i$ associated with the $\mathbb{R}^{d-1}$ translation symmetry and the remaining coordinate to be some parameter~$\eta$ along the null direction.  The parameter $\eta$ is arbitrary and need not be affine; indeed, for a non-trivial null shell the affine parameter is discontinuous across the shell and one cannot take $\eta$ to be affine on both sides.

We also introduce tangent vectors
\be
e^\alpha_i \equiv \frac{\partial X^\alpha}{\partial x^i}, \quad k^\alpha \equiv e^\alpha_\eta \equiv \frac{\partial X^\alpha}{\partial \eta},
\ee
and an auxiliary null vector~$N^\alpha$ which satisfies~$N_\alpha k^\alpha = -1$.  Note that both~$k^\alpha$ and~$N^\alpha$ are orthogonal to the transverse tangent vectors~$e^\alpha_i$.

The relevant results from~\cite{Poisson} are as follows.  The induced metric on a shell is
\be
\sigma_{ij} = g_{\alpha\beta}\left(X^\alpha\right) e^\alpha_i e^\beta_j,
\ee
which for regularity is required to be the same when calculated from either side of a given shell.  The transverse extrinsic curvature of a shell is
\be
C_{ab} \equiv -N_\alpha e^\beta_a \grad_\beta e^\alpha_b,
\ee
which need not be the same on the two sides.  The difference in transverse curvature across the shell gives the shell stress tensor.  It is convenient to decompose this tensor into a surface energy density~$\mu$, energy current~$j^i$, and pressure~$p$:
\be
\mu = -\frac{1}{8\pi G_N} \sigma^{ij} \left(C_{ij}^+ - C_{ij}^-\right), \quad j^i = \frac{1}{8\pi G_N} \sigma^{ij} \left(C_{j \eta}^+ - C_{j \eta}^-\right), \quad p = -\frac{1}{8\pi G_N} \left(C_{\eta \eta}^+ - C_{\eta \eta}^-\right),
\ee
where the~$+ (-)$ superscripts imply that the quantity is calculated on the side of the shell into (away from) which~$k^\alpha$ points, and~$G$ is the full~$(d+1)$-dimensional Newton's constant.

Since our shells lie on horizons~$r = r_+$, the induced metric on each shell is just
\be
ds^2_\mathrm{shell} = r_+^2 \, d\vec{x}_{d-1}^2.
\ee
To construct the single-shell spacetime of figure~\ref{fig:cutpaste} with the edges~$\E$ of~$\mathscr{I}_\mathrm{dS}$ at advanced time~$v_I = v_0$, we use the embeddings
\be
r = r_+, \quad v_{I} = \eta_A, \quad v_{II} = \frac{1}{\kappa_{II}} \, g_A\left(\kappa_{I}\eta_A\right),
\ee
where~$\kappa_n \equiv |f'_n(r_+)|/2$ are the surface gravities of each horizon,~$\eta_A$ is a parameter along the generators of the shell, and as stated in the main text~$g_A(x)$ is an arbitrary continuous and monotonically increasing function with range~$(-\infty,\infty)$ and domain~$(v_0,\infty)$.  The density, current, and pressure of shell~$A$ are then
\begin{subequations}
\label{eq:BT}
\bea
\mu_A &= \frac{d-1}{8\pi G_N r_+}\left[\frac{\kappa_{II}}{\kappa_I g_A'(\kappa_I \, \eta_A)} - 1\right], \\
j^i_A &= 0, \\
p_A &= \frac{\kappa_I}{8\pi G_N}\left[1 + g_A'(\kappa_I\eta_A) - \frac{g_A''(\kappa_I\eta_A)}{g_A'(\kappa_I\eta_A)}\right].
\eea
\end{subequations}

\begin{figure}[t]
\centering
%
%
%
\includegraphics[page=12]{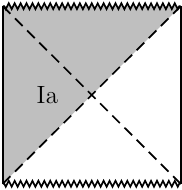}
\hspace{1cm}
%
%
%
\includegraphics[page=13]{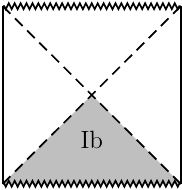}
\hspace{1cm}
%
%
%
\includegraphics[page=14]{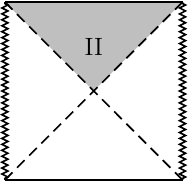}
\caption{Conformal diagrams from which we cut our (shaded) regions~Ia,~Ib, and~II.}
\label{fig:patches}
\end{figure}

To instead construct the doubly-patched spacetime shown in the lower panel of figure~\ref{fig:Vaidyapatching}, we leave patches~Ib and~II and shell~$A$ untouched (that is, patch~Ib is just the corresponding piece of the original patch~I with~$\ell_{Ib} = \ell_I$), and we take patch~Ia to be the exterior of a Schwarzschild-AdS black hole with horizon size~$r_+$ and AdS radius~$\ell_{Ia}$.  The three patches we stitch together are shown in figure~\ref{fig:patches}, and their metrics are as in~\eqref{eq:patchmetric} with
\begin{subequations}
\bea
f_{Ia}(r) &= \frac{r^2}{\ell_{Ia}^2}\left(1-\left(\frac{r_+}{r}\right)^d\right), \\
f_{Ib}(r) &= -\frac{r^2}{\ell_{Ib}^2}\left(\left(\frac{r_+}{r}\right)^d-1\right), \\
f_{II}(r) &= -\frac{r^2}{\ell_{II}^2}\left(1-\left(\frac{r_+}{r}\right)^d\right).
\eea
\end{subequations}

In terms of the Eddington-Finkelstein coordinates~\eqref{eq:EF}, the embeddings of shell~$B$ in patches~Ia and~Ib are
\be
r = r_+, \quad u_{Ia} = -\frac{1}{\kappa_{Ia}} \, \ln\left(-\kappa_{Ia} \eta_B\right), \quad u_{Ib} = \frac{1}{\kappa_{Ib}} \, g_B\left(\kappa_{Ia} \eta_B\right),
\ee
where as for shell~$A$,~$g_B(x)$ is an arbitrary continuous and monotonically increasing function that maps~$(-\infty,\infty) \mapsto (-\infty,\infty)$.  Note that with this embedding,~$\eta_B$ is an affine parameter along the shell with respect to the metric of patch~Ia.  The density, current, and pressure of shell~$B$ are then
\begin{subequations}
\bea
\label{eq:AT}
\mu_B &= \frac{d-1}{8\pi G_N r_+}\left[\kappa_{Ia} \eta_B + \frac{\kappa_{Ib}}{\kappa_{Ia} \, g_B'(\kappa_{Ia}\eta_B)}\right], \\
j^i_B &= 0, \\
p_B &= \frac{\kappa_{Ia}}{8\pi G_N}\left[ \frac{g_B''(\kappa_{Ia}\eta_B)}{g_B'(\kappa_{Ia}\eta_B)} - g_B'(\kappa_{Ia}\eta_B)\right].
\eea
\end{subequations}
Note that~$p_B$ vanishes only if~$g_B(x) = \mathrm{const.}$ or~$g_B(x) = -\ln(x+c)$, neither of which is compatible with the continuity and monotonicity of~$g_B$.  So as claimed in footnote~\ref{note:pressure}, shell~$B$ cannot be pressureless, and this spacetime is not a limiting case of AdS-Vaidya.

Nevertheless, the null energy condition can be satisfied for an appropriate choice of parameters.  Indeed, for any~$d \geq 2$, let
\be
\label{eq:Aparams}
\quad \kappa_{Ia} r_+ = 1, \quad \kappa_{Ib} r_+ = 1, \quad g_B(x) = \mathrm{arcsinh}(x).
\ee
Then we find that
\be
\mu_B + p_B = \frac{d}{8\pi G_N r_+}P(\kappa_{Ia}\eta_B)\left(1 - \frac{\Delta(\kappa_{Ia} \eta_B)}{d}\right),
\ee
where
\be
P(x) = x+\sqrt{1+x^2} \quad \mbox{and} \quad \Delta(x) = 1+\frac{1}{1+x^2}
\ee
satisfy~$P(x) > 0$ and~$\Delta(x) \leq 2$ everywhere.  It then follows that~$\mu_B + p_B \geq 0$ for all~$d \geq 2$.

\section{Regulated Wormholes}
\label{app:regulated}

This appendix considers simple models of the regulated wormholes mentioned in sections \ref{sec:theorem} and
\ref{sec:total} in which inflation ends on a finite surface, after which the wormhole collapses to a singularity.  The simplification made here is that sections \ref{sec:theorem} and \ref{sec:total} required this singularity to be everywhere of Kasner or of big crunch type (see footnote~\ref{sing}), but the examples below will violate this condition at the regulated analogues of the edges~$\E$ of~$\mathscr{I}_\mathrm{dS}$.  The point is that it is convenient to retain symmetry of patch~II under the Killing field~$\xi$ of section~\ref{sec:cutpaste}.  But since the orbits of~$\xi$ approach~$\E$, this means that surfaces of constant scale factor will also approach~$\E$ in the regulated spacetimes.  The singularity of our regulated spacetimes thus fails to be either Kasner-like or of big crunch type at~$\E$.

\begin{figure}[t]
\centering
%
%
%
\includegraphics[page=15]{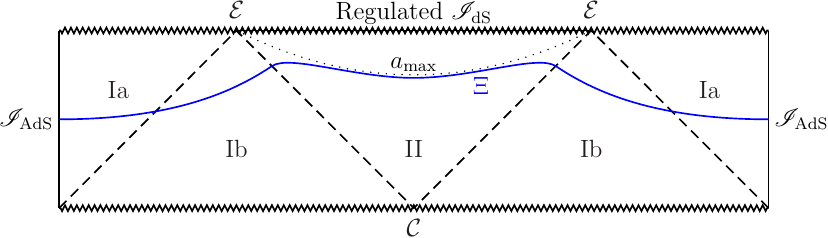}
\caption{A regulated AdS-dS-wormhole with a finite amount of inflation followed by collapse to a singularity.  The dotted line labeled~$a_\mathrm{max}$ indicates the surface on which the effective scale factor \eqref{eq:aeff} in patch~II reaches a maximum; this slice serves as an accumulation surface for wormhole-spanning extremal surfaces.  In the cut-and-paste geometry, the proper distance between any point in patch~II and either of the boundary points~$\E$ is infinite; this is an artefact of the cut-and-paste construction, and will no longer be true for appropriately smoothed out null shells.  Such smoothed cases lead to the existence of finite-area HHRT surfaces such as the one labeled~$\Xi$ (solid line, blue in color version).}
\label{fig:regulated}
\end{figure}

Retaining symmetry along~$\xi$ takes the above singularity to lie at a proper time~$\tau$ along the worldline of any freely falling observer chosen to start at $\tau =0$ from the point labeled~$\mathcal{C}$ on the past boundary of figure~\ref{fig:regulated}.  In the limit~$\tau \to \infty$, we recover the original AdS-dS-wormhole.

Such regulated wormholes can be constructed as in Section~\ref{sec:cutpaste} above by replacing the metric in patch~II with
\be
\label{eq:regmetric}
ds^2 = -d\rho^2 + R^2(\rho) dt^2 + X^2(\rho) d\vec{x}_{d-1}^2,
\ee
where~$\rho \in [0,\tau)$ is the proper time along worldlines of freely falling observers with constant~$t,x^i$.  Near~$\rho = 0$, we impose that~$X = r_+ + \cdots$ and~$R = \kappa_{II} \rho + \cdots$ where $\cdots$ represent terms that vanish as $\rho \rightarrow 0$.  Then to good approximation~$\rho = 0$ remains a horizon with surface gravity~$\kappa_{II}$, and in particular the regulated spacetime~\eqref{eq:regmetric} can be patched into the wormhole using the same null shells (with precisely the same stress tensor) as in section~\ref{sec:cutpaste}.  In these coordinates, the patch~II metric~\eqref{eq:patchmetric} of the original unregulated spacetime corresponds to
\be
\label{eq:RXunreg}
R(\rho) = \frac{r_+}{\ell} \tanh\left(\frac{d\rho}{2\ell}\right) \cosh^{2/d}\left(\frac{d\rho}{2\ell}\right), \quad X(\rho) = r_+\cosh^{2/d}\left(\frac{d\rho}{2\ell}\right).
\ee

It is straightforward to identify extremal surfaces for which $t$ and $d-2$ of the $\vec x_{d-1}$ are constant.  These are the analogue in region II of surfaces found by Hartman and Maldacena \cite{Hartman:2013qma} to be attractors for more generic extremal surfaces in the two-sided planar AdS-Schwarzschild black hole in the limit where both boundaries of the extremal surface are anchored to very late times on the two AdS boundaries.  The area of our highly symmetric surfaces is governed in region II by the effective scale factor
\be
\label{eq:aeff}
a_\mathrm{eff}(\rho) = R(\rho) X^{d-2}(\rho),
\ee
and extremal such surfaces lie at extrema of $a_\mathrm{eff}(\rho)$.  In parallel with \cite{Hartman:2013qma}, in the fully-regulated case we expect one of these extrema to be a late-time attractor with the actual wormhole-spanning extremal surface staying very close to one of these surfaces across most of region II, as shown in figure~\ref{fig:regulated}.  The maximin argument of section \ref{sec:theorem} suggests that the desired extremal surfaces in fact accumulate along the global maximum of $a_\mathrm{eff}(\rho) = a_\mathrm{max}$.  In fact, note that our cut-and-paste construction renders the area of the attractor surface~$a_\mathrm{max}$ infinite, since any point in patch~II is an infinite distance from either of the boundary points marked~$\E$.  As a result, the wormhole-spanning extremal surfaces in this geometry still have infinite area.  However, it is clear that this is simply an artefact of our patching procedure, which causes~$\E$ to violate the conditions of footnote~\ref{sing}.  By smoothing out the null shells, the distance to any~$\E$ from patch~II becomes finite, and thus so does the area of the~$a_\mathrm{max}$.  These smoothed-out regulated AdS-dS-wormholes thus have HHRT surfaces with finite areas that grow without bound as we increase $a_\mathrm{max}$.

We now construct explicit examples of the above (unsmoothed) regulated wormholes and verify the above conjecture concerning wormhole-spanning extremal surfaces.  To do so we couple gravity to a scalar field $\phi$, so that the action is
\be
\label{eq:action}
S = \frac{1}{16\pi G_N} \int d^d x \sqrt{-g} \, R - \int d^d x \sqrt{-g} \left(\frac{1}{2} \, g^{\mu\nu} \grad_\mu \phi \grad_\nu \phi + V(\phi)\right).
\ee
We set $\phi = \phi(\rho)$ and take the metric to be \eqref{eq:regmetric}, in which the coordinate~$\rho$ plays the role of a proper time.  The equations of motion obtained from the action~\eqref{eq:action} can be rearranged into
\begin{subequations}
\label{eqs:EOM}
\bea
\frac{2X'R'}{XR} + (d-2)\left(\frac{X'}{X}\right)^2 - \frac{8\pi G_N}{d-1}\left(\left(\phi'\right)^2 + 2V(\phi)\right) &= 0, \label{subeq:constraint} \\
\frac{X''}{X} - \frac{X'R'}{XR} + \frac{8\pi G_N}{d-1}\left(\phi'\right)^2 &= 0, \\
\frac{R''}{R} + (d-3)\frac{X'R'}{XR} - (d-2)\left(\frac{X'}{X}\right)^2 + \frac{8\pi G_N}{d-1}\left(\phi'\right)^2 &= 0, \\
\phi'' + \left(\frac{R'}{R}+(d-1)\frac{X'}{X}\right)\phi' + V'(\phi) &= 0.
\eea
\end{subequations}
Note that~\eqref{subeq:constraint} is a constraint equation, while the other three are dynamical.  As usual, the constraint is conserved by the dynamical equations, so that there are only three independent equations that must be solved.

Solutions to~\eqref{eqs:EOM} will be characterized by some~$\rho$ at which~$X$,~$R$, and~$\phi$ become singular; without loss of generality we take this time to be~$\rho = 0$.  Then one can show that for polynomial~$V(\phi)$, the solutions near such singular points behave like
\begin{subequations}
\label{eqs:expansions}
\bea
X(\rho) &= \rho^r \left[ X_{00} + \sum_{n=1}^\infty \sum_{m=0}^{nN} X_{n,m} \rho^{2n}  \left(\ln \rho/\rho_0 \right)^m \right], \\
R(\rho) &= \rho^{1-3r} \left[ R_{00} + \sum_{n=1}^\infty \sum_{m=0}^{nN} R_{n,m} \rho^{2n}  \left(\ln \rho/\rho_0 \right)^m \right], \\
\phi(\rho) &= \phi_{00} + \phi_{01} \ln \rho/\rho_0 + \sum_{n=1}^\infty \sum_{m=0}^{nN} \phi_{n,m} \rho^{2n}  \left(\ln \rho/\rho_0 \right)^m,
\eea
\end{subequations}
where~$\rho_0$ is some arbitrary scale, the integer~$N$ is the highest power of~$\phi$ appearing in~$V(\phi)$, and~$r$,~$X_{00}$,~$R_{00}$,~$\phi_{00}$, and~$\phi_{01}$ are free parameters subject to the constraint~$8\pi G_N \phi_{01}^2 = 6r(1-2r)$.

The near-horizon behavior requires~$r = 0$ (and therefore~$\phi_{01} = 0$) as well as~$X_{00} = r_+$ and~$R_{00} = \kappa_{II}$.  The condition~$\phi_{01} = 0$ can be interpreted as the statement that the energy density of the scalar field must be finite at the horizon, or else backreaction would destroy the near-horizon geometry.  Furthermore,~$r = 0$ implies that~$\phi'(0) = 0$, so that the scalar field starts at rest at the horizon and evolves according to the form of~$V(\phi)$.

\begin{figure}[t]
\centering
%
%
%
%
%
%
\includegraphics[page=16]{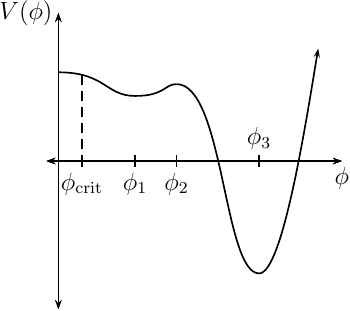}
\caption{A sketch of the potential we consider.  At the horizon, the scalar field is at rest at some~$\phi_0$ and is then allowed to roll down the potential.  If~$\phi_0 < \phi_\mathrm{crit}$, the scalar field falls into the minimum at~$\phi_3$; if~$\phi_0 = \phi_\mathrm{crit}$, the scalar field stops at~$\phi_2$, and if~$\phi_\mathrm{crit} < \phi_0 < \phi_2$, the scalar field falls into the minimum at~$\phi_1$.}
\label{fig:potential}
\end{figure}

By choosing $V(\phi) = \mathrm{const.} > 0$ and~$\phi'(0) = 0$, we obtain the unregulated solution~\eqref{eq:RXunreg}.  In order to obtain a regulated solution that crunches in finite proper time, we require a potential~$V(\phi)$ with extrema at both~$V(\phi) > 0$ and~$V(\phi) < 0$.  We therefore consider a potential of the form shown in figure~\ref{fig:potential}; explicitly, we take
\be
\label{eq:potential}
V(\phi) = h^2\left[\frac{1}{20} - \frac{3}{16} \left(\frac{\phi}{\phi_*}\right)^2 + \frac{7}{5} \left(\frac{\phi}{\phi_*}\right)^4 - 4\left(\frac{\phi}{\phi_*}\right)^6 + 3\left(\frac{\phi}{\phi_*}\right)^8\right],
\ee
where~$h$ is an overall scale that sets the height of the potential and~$\phi_*$ is a reference scale.  This potential has local maxima at the origin and some~$\phi_2$, and local minima at some~$\phi_1$ and~$\phi_3$.  In particular, it satisfies~$V(0) > 0$,~$V(\phi_1) > 0$,~$V(\phi_2) > 0$, but~$V(\phi_3) < 0$.

To construct a solution, the scalar field is released at some initial value~$\phi_0$ at which~$V(\phi_0)>0$.  If~$\phi_0$ is smaller than some critical value~$\phi_\mathrm{crit}$, the scalar field rolls past the extrema~$\phi_1$ and~$\phi_2$ and into the AdS extremum~$\phi_3$, where~$V(\phi_3) < 0$.  This produces a negative effective cosmological constant, causing the solution to become singular in finite~$\rho$.  As~$\phi_0$ is increased closer to~$\phi_\mathrm{crit}$, the scalar field spends more and more time near the maximum~$\phi_2$, yielding a spacetime with a longer and longer expanding region before the singularity.  Eventually, when~$\phi_0 = \phi_\mathrm{crit}$, the initial conditions are tuned such that the scalar field remains at~$\phi_2$ indefinitely, yielding a version of the unregulated AdS-dS-wormhole\footnote{For~$\phi_\mathrm{crit} < \phi_0 < \phi_2$, the scalar field comes to rest at~$\phi_1$, again producing an unregulated AdS-dS-wormhole.}.  Thus the regulator~$\tau$ can be made arbitrarily large by taking~$\phi_0$ arbitrarily close to~$\phi_\mathrm{crit}$.

\begin{figure}[t]
\centering
\includegraphics[width=0.4\textwidth]{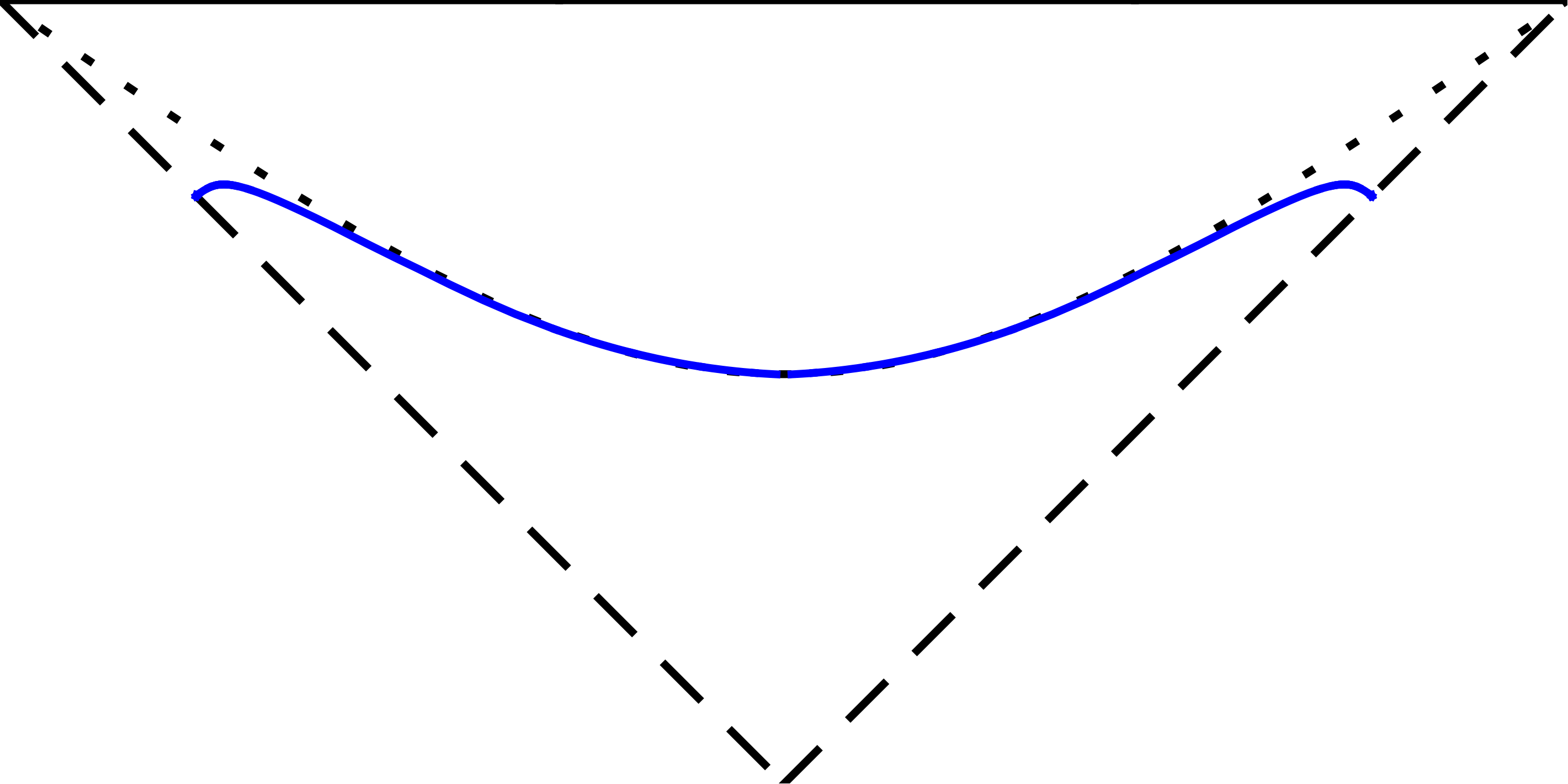} \quad
\includegraphics[width=0.4\textwidth]{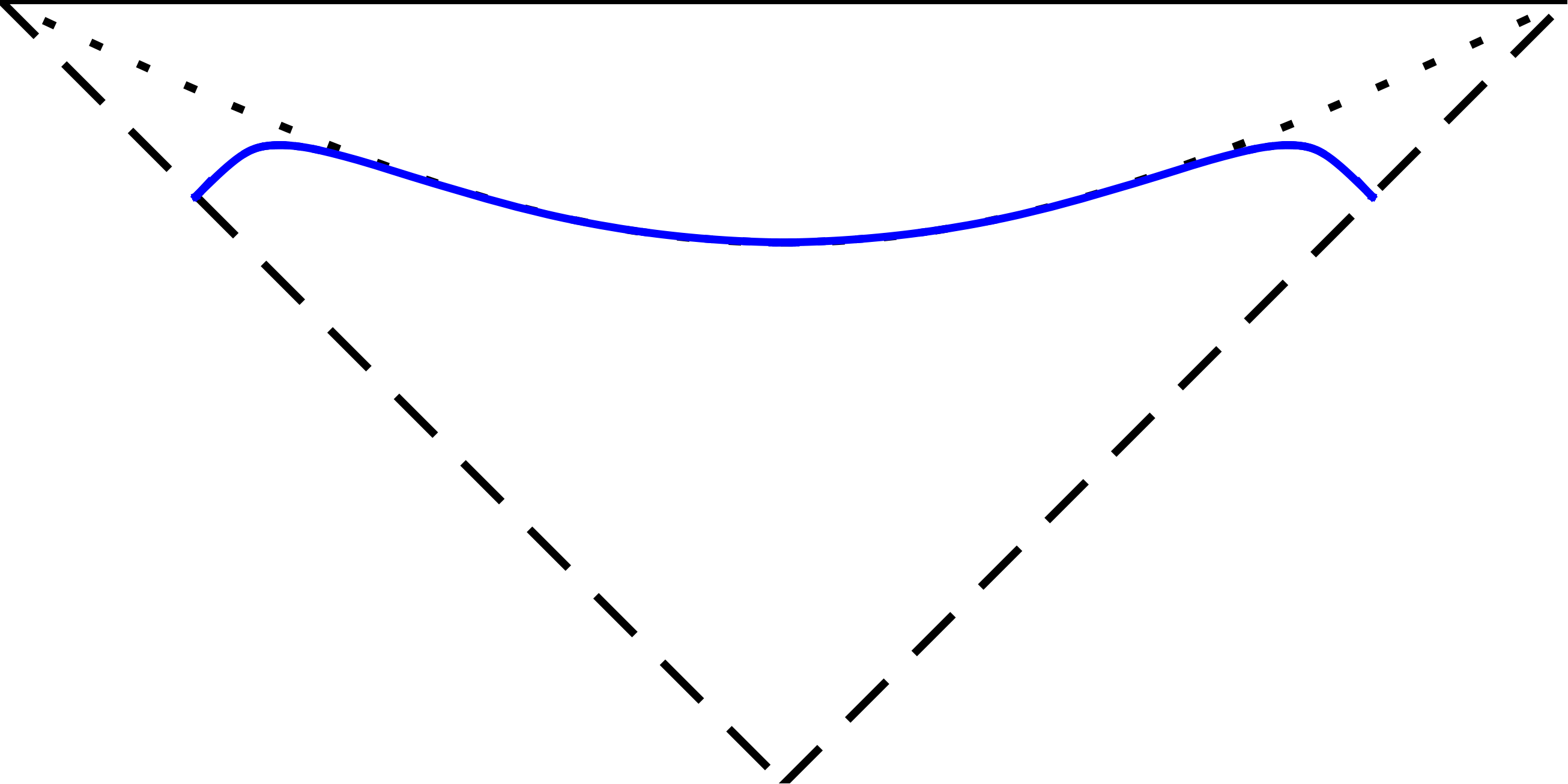} \\
\vspace{0.5cm}
\includegraphics[width=0.4\textwidth]{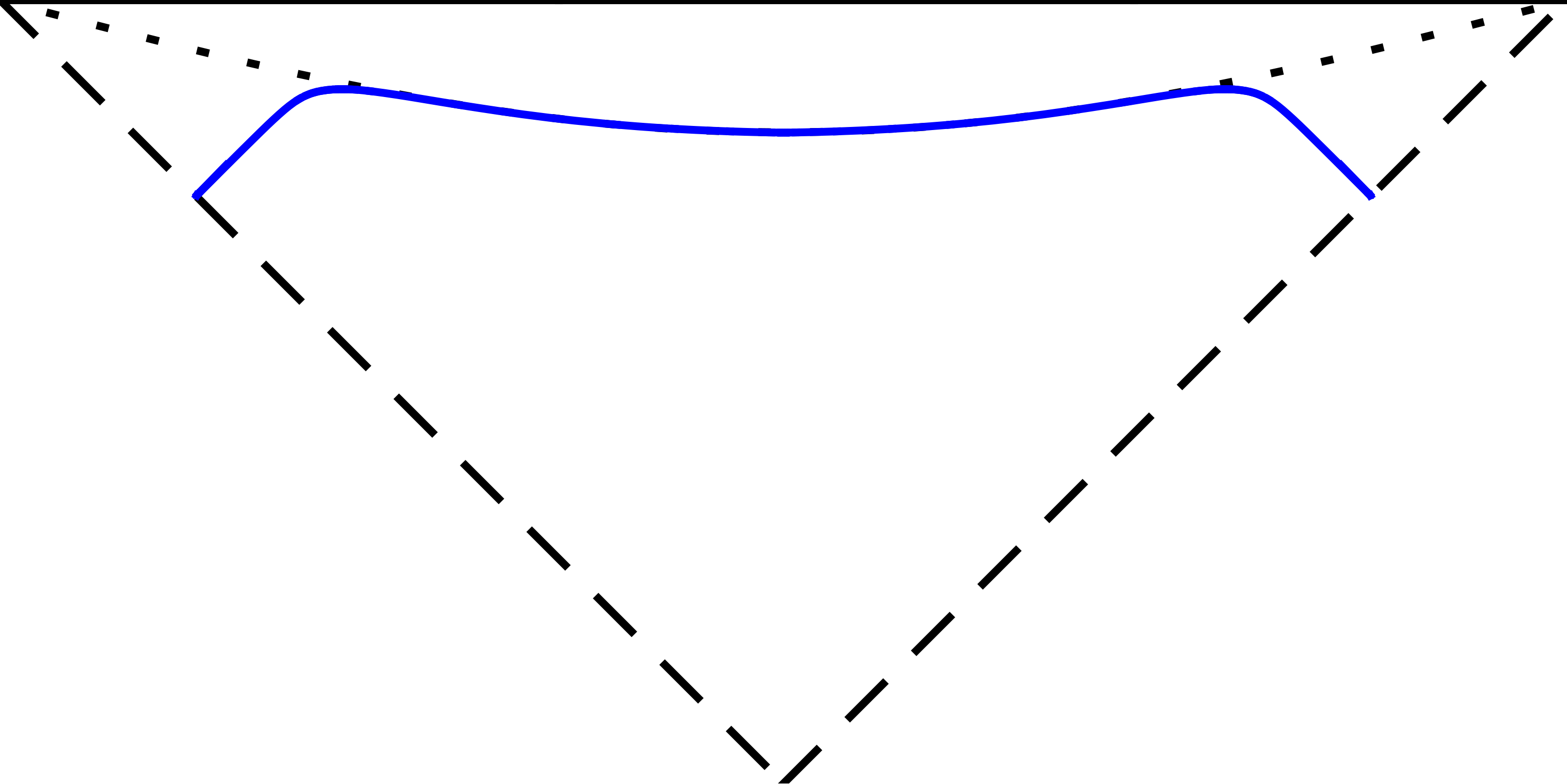} \quad
\includegraphics[width=0.4\textwidth]{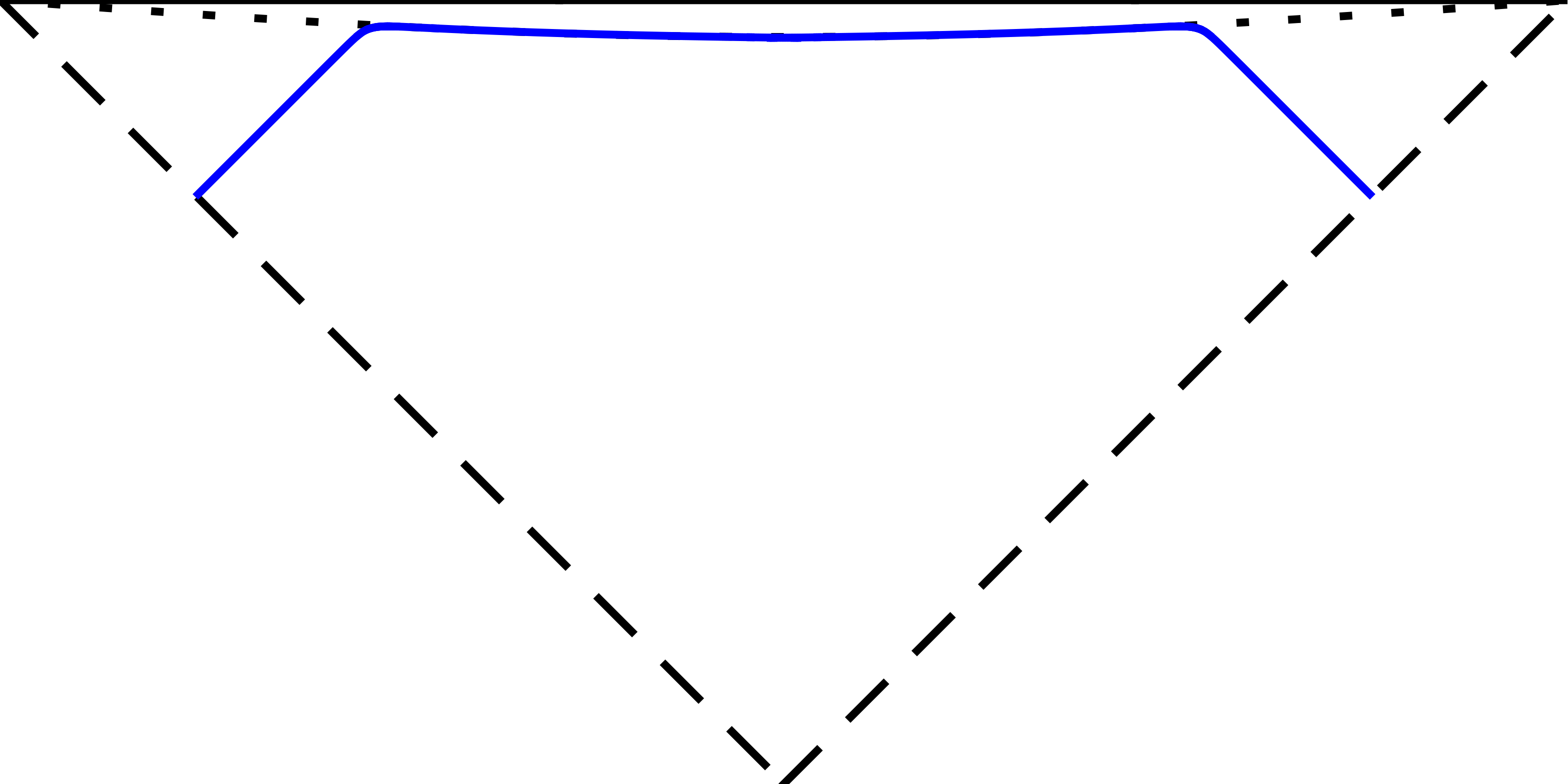}
\caption{The regulated asymptotically~dS patch for $d=4$ and various values of~$\phi_0$; from left to right, top to bottom, the figures have~$(\phi_\mathrm{crit} - \phi_0)/\phi_* = 10^{-1}$,~$10^{-2}$,~$10^{-3}$,~$10^{-5}$, corresponding to~$\sqrt{8\pi G_N} \, h \tau \approx 2.5$,~$3.2$,~$3.9$,~$5.5$.  The dotted lines mark the maxima~$a_\mathrm{max}$ of~$a_\mathrm{eff}$, while the solid curves (blue in color version) show extremal surfaces that enter through the horizon.  The solid horizontal lines are singularities, which the extremal surfaces are prevented from reaching.  For~$\phi_0 = \phi_\mathrm{crit}$,~$\tau = \infty$ and~$a_\mathrm{max}$ merges with the singularity to create future dS infinity~$\mathscr{I}_\mathrm{dS}$.  The extremal surfaces then cease to exist in the Lorentzian section.}
\label{fig:regulatedsurfaces}
\end{figure}

Finally we consider wormhole-spanning extremal surfaces in smoothed, regulated wormholes that satisfy the conditions of footnote~\ref{sing} everywhere.  Note that any wormhole-spanning surface~$\Xi$ must pass through patch II, entering and leaving this patch through the de Sitter horizon $\rho=0$.  For our unsmoothed cut-and-paste geometries,~$\Xi$ will cross the de Sitter horizon in the far future in order to run along the entire (infinite) length of the accumulation surface~$a_\mathrm{max}$.  Smoothing out the null shells to obtain a finite-area~$\Xi$ will keep these anchors at a finite place.  The exact point of crossing is determined by balancing the tendency to maximize the area in patch~I (which tends to flatten~$\Xi$ in this region) with the tendency to run along the~$a_\mathrm{max}$ surface in patch~II.  So as the anchors on~$\mathscr{I}_\mathrm{AdS}$ move to the far future, so does the intersection of~$\Xi$ with the~dS horizon.  It is thus sufficient to study codimension-2 extremal surfaces anchored at $\rho=0$ in the limit where these anchors are taken to the far future.  Sample such surfaces are plotted numerically\footnote{These solutions were found by integrating the equations of motion~\eqref{eqs:EOM} using~\texttt{Mathematica}'s built-in~\texttt{NDSolve} command, which is more than sufficient for generating the desired figures.} in figure~\ref{fig:regulatedsurfaces} for $d=4$ in comparison with surfaces on which~$a_\mathrm{eff}(\rho)$ (defined in~\eqref{eq:aeff}) attains its maximum.   We find~$\phi_\mathrm{crit} \approx 0.21 \, \phi_*$.

\section{Correlators in dS$_3$}
\label{app:2pt}

We now show how the geodesic approximation in~dS$_3$ reproduces the large-mass behavior of the Wightman function
of a free massive scalar field in the Hadamard de Sitter-invariant (Bunch-Davies) vacuum.  As is well known (see e.g. \cite{Schomblond:1976xc,Bunch:1978yq} for $d=1,3$), for~dS$_{d+1}$ this two-point function is
\be
\label{eq:dSwightman}
G(x,x') = \frac{H^{d-1}}{(4\pi)^{(d+1)/2}} \, \frac{\Gamma(-c)\Gamma(c+d)}{\Gamma((d+1)/2)} \, _2F_1\left(-c,c+d; \frac{d+1}{2} ; \frac{1+Z}{2}\right),
\ee
where
\be
c = -\frac{d}{2} + \sqrt{\frac{d^2}{4} - \frac{m^2}{H^2}},
\ee
$H$ is the Hubble constant,~$m$ is the mass of~$\phi$, and~$Z$ is the de Sitter invariant given by the inner product of unit vectors associated with the standard embedding of~dS$_{d+1}$ into $d+2$ Minkowski space.   In the coordinates of~\eqref{eq:deSitterflat} we have
\be
Z(x,x') = 1 + \frac{(e^{-Ht} - e^{-Ht'})^2 - H^2(\vec{x} - \vec{x}')^2}{2e^{-Ht} e^{-Ht'}}.
\ee

We wish to study~\eqref{eq:dSwightman} for~$d = 2$ and~$m/H \gg 1$.  Using the identities~\cite{Prudnikov:1986}
\begin{subequations}
\be
\Gamma(1+z)\Gamma(1-z) = \frac{\pi z}{\sin(\pi z)},
\ee
\begin{multline}
_2F_1\left(a, 2-a; \frac{3}{2}; -z^2\right) = \\ \frac{1}{4(a-1)z\sqrt{1+z^2}}\left[\left(\sqrt{1+z^2}+z\right)^{2(a-1)} - \left(\sqrt{1+z^2}-z\right)^{2(a-1)}\right],
\end{multline}
\end{subequations}
for~$d = 2$ we find
\be
G(t_0, \vec{x}; t_0, \vec{x}') = \frac{H}{4\pi} \, \frac{1}{\widetilde{L} \sin(\pi \nu)} \begin{cases} \frac{1}{\sqrt{1-\widetilde{L}^2/4}} \sin(2\nu \arccos(\widetilde{L}/2)), & \widetilde{L} < 2, \\
\frac{1}{\sqrt{\widetilde{L}^2/4-1}} \sinh(2\nu \, \arccosh(\widetilde{L}/2)), & \widetilde{L} > 2, \end{cases}
\ee
where~$\nu \equiv \sqrt{1-m^2/H^2}$ and~$\widetilde{L} \equiv H e^{H t_0} |\vec{x} - \vec{x}'|$.  For large~$m/H$ we have~$\nu \approx im/H$ and thus
\be
\label{eq:correlator}
G(t_0, \vec{x}; t_0, \vec{x}') \sim \frac{e^{-\pi m/H}}{1-e^{-2\pi m/H}} \begin{cases} e^{(2m/H)\arccos(\widetilde{L}/2)} - e^{-(2m/H)\arccos(\widetilde{L}/2)}, & \widetilde{L} < 2, \\
e^{(2im/H)\arccosh(\widetilde{L}/2)} - e^{-(2im/H)\arccosh(\widetilde{L}/2)}, & \widetilde{L} > 2, \end{cases}
\ee
where the $\sim$ indicates that we have dropped polynomial corrections to exponentials in~$m$; i.e., we have kept terms that in a saddle point approximation can come from a sum over saddles.  The remaining terms may well come from fluctuations around these saddles, though we will not consider this in detail.  Note that since the factor~$1-e^{2\pi m/H}$ lies in the denominator of~\eqref{eq:correlator}, it in fact leads to an infinite number of terms exponential in~$m$.

We now make explicit that spacelike geodesics can reproduce the exponential terms in~\eqref{eq:correlator}.  For~$d = 2$, expressions~\eqref{eq:dSL} and~\eqref{eq:dSarea} simplify to
\begin{subequations}
\bea
\widetilde{L} &= 2e^{-H \Delta t} \sqrt{e^{2H\Delta t}-1}, \\
A^\pm_n &= \frac{2}{H} \left[\pm \arctan\sqrt{e^{2H \Delta t} - 1} + n\pi\right],
\eea
\end{subequations}
where now~$\widetilde{L} = 2 H e^{H t_0} L$.  The~$\pm$ sign and the integer~$n$ that appear in~$A_n^\pm$ represent the analytic continuation to all sheets of the square root and inverse tangent, respectively.  Writing~$A_n^\pm$ in the form
\be
A_n^\pm = \frac{1}{H} \begin{cases} (2n\pm 1)\pi \mp 2\arccos\left(\widetilde{L}/2\right), & \widetilde{L} < 2, \\
(2n\pm 1)\pi \mp 2i \, \arccosh\left(\widetilde{L}/2\right), & \widetilde{L} > 2, \end{cases}
\ee
one may interpret each term as the length of a distinct (possibly complex) geodesic. Comparing with the exact expression~\eqref{eq:correlator} shows that
\be
G(t_0, \vec{x}; t_0, \vec{x}') \sim \sum_{2n\pm 1 \geq -1} c_n^\pm e^{-mA_n^\pm}
\ee
for appropriate order-1 phases~$c_n^\pm$ (which in the saddle-point approximation are higher order effects determined by fluctuations around each saddle).  Since the sum is over precisely those $n$ and signs $\pm$ with $2n\pm 1 \geq -1$, we conclude that these are the saddles that contribute to the desired path integral.  It is interesting that this represents a sum over both all sheets in the Riemann surface for $\widetilde L(\Delta t)$ and an infinite number of sheets in the Riemann surface for $A(\widetilde L)$, though sufficiently ``negative'' sheets are not included.  We note that $d=2$ is a special case where $A(\widetilde L)$ (understood as a map from the Riemann surface for~$\widetilde L(\Delta t)$ to the Riemann surface for~$A(\Delta t)$) is multi-valued; in higher dimensions we expect that as in \cite{Fischetti:2014zja} one can take $A(\widetilde L)$ to be single valued, since the Riemann surface for $\widetilde L(\Delta t)$ has an infinite number of sheets for $d > 2$.


\bibliographystyle{jhep}
\bibliography{biblio}

\end{document}